\documentclass[journal,letterpaper,10pt]{IEEEtran}
\IEEEoverridecommandlockouts
\usepackage{cite}
\usepackage{amsmath,amssymb,amsfonts,amsthm}
\usepackage{graphicx}
\usepackage{textcomp}
\usepackage{xcolor}
\usepackage{dsfont}
\usepackage{bm}
\usepackage[left=0.625in, right=0.625in, top=0.7in, bottom=1.0in]{geometry}

\usepackage{booktabs} 

\usepackage{tikz}
\usetikzlibrary{arrows.meta,positioning,shapes,calc}

\usepackage{subcaption}

\usepackage{algorithm}
\usepackage[noend]{algorithmic}

\newtheorem{theorem}{Theorem}
\newtheorem{lemma}{Lemma}
\newtheorem{corollary}{Corollary}
\newtheorem{definition}{Definition}
\newtheorem{proposition}{Proposition}
\newtheorem{remark}{Remark}

 \title{Controlling Activity Leakage in Wi-Fi CSI Feedback: A Systematic Design Approach} 

\author{
    \IEEEauthorblockN{
        Mohamed Seif$^{1}$,
        Atsutse Kludze$^{1}$,
        Yasaman Ghasempour$^{1}$,
        H.~Vincent Poor$^{1}$,
        Doru Calin$^{2}$,
        Andrea J.~Goldsmith$^{3}$
    }
    \IEEEauthorblockA{$^{1}$Princeton University}
    \IEEEauthorblockA{$^{2}$MediaTek}
    \IEEEauthorblockA{$^{3}$Stony Brook University}
\thanks{}   
}

\author{
    \IEEEauthorblockN{
        Mohamed Seif$^{1}$,
        Atsutse Kludze$^{1}$,
        Yasaman Ghasempour$^{1}$,
        H.~Vincent Poor$^{1}$,
        Doru Calin$^{2}$,
        Andrea J.~Goldsmith$^{3}$
    }
    \IEEEauthorblockA{$^{1}$Princeton University}
    \IEEEauthorblockA{$^{2}$MediaTek}
    \IEEEauthorblockA{$^{3}$Stony Brook University}
    
\thanks{This work was supported by the AFOSR award \#002484665, the U.S. National Science Foundation under Grants: CNS-2148271, CNS-2147631, and an Innovation Grant from Princeton NextG. Part of this work was presented at the 2026 IEEE International Symposium on Dynamic Spectrum Access Networks (DySPAN).}

}

\date{}

\begin{document}
\maketitle

\begin{abstract}
Explicit channel state information  (CSI) feedback in IEEE~802.11 conveys  \emph{transmit beamforming directions} by reporting quantized Givens rotation and phase angles that parametrize the right-singular subspace of the channel matrix. Because these angles encode fine-grained spatial signatures of the propagation environment, recent work have shown that plaintext CSI feedback can inadvertently reveal user activity, identity, and location to passive eavesdroppers. In this work, we introduce a standards-compatible \emph{differentially private (DP)  quantization mechanism} that replaces deterministic angular quantization with a stochastic quantizer with a localized differential privacy guarantee, applied directly to the Givens parameters of the transmit beamforming matrix. The mechanism preserves the 802.11 feedback structure, admits closed-form sensitivity bounds for the angular representation, and enables principled privacy calibration. Numerical simulations demonstrate strong privacy guarantees with minimal degradation in beamforming performance.
\end{abstract}


\section{Introduction}
\label{sec:intro}

Beamforming is a core capability of modern multi-antenna wireless systems \cite{goldsmith2005wireless, geraci2025wi}, enabling transmitters to dynamically shape and steer radiated energy toward intended receivers. By coherently combining signals across multiple antennas, the transmitter forms highly directional beams that substantially improve the received signal-to-noise ratio (SNR) while suppressing unintended interference—capabilities unattainable with omnidirectional transmission. Introduced into Wi-Fi through the IEEE~802.11n “High Throughput’’ amendment \cite{ieee80211n2012}, beamforming has since become standard in commodity WLAN devices, where the access point (beamformer) computes a channel-dependent steering matrix that optimizes reception quality at the client (beamformee).

Two primary approaches are used to compute this steering matrix \cite{perahia2013next}.
In \emph{implicit} beamforming, the transmitter estimates the downlink channel from uplink pilot transmissions under an assumption of channel reciprocity. While attractive for its simplicity, this approach is often suboptimal in practice due to hardware asymmetries between the transmit and receive chains. By contrast, \emph{explicit} beamforming, now the mandated method in modern Wi-Fi standards, achieves higher accuracy by relying on channel state information (CSI) fed back by the client. The beamformee estimates the downlink channel, compresses it, and reports the result to the beamformer through dedicated sounding and feedback frames, typically using quantized matrices or precoding matrix indicators (PMIs).

\subsection{Privacy Concerns in Explicit Beamforming}

Although \emph{explicit} beamforming enables fine-grained directional control and high throughput, it also introduces a previously underappreciated \emph{privacy attack surface}: the CSI feedback itself encodes detailed spatial and motion characteristics of users and their surrounding environment. In this work, we revisit CSI feedback not only as a mechanism for link optimization, but also as a potential channel for privacy leakage, and we develop a provably private beamforming feedback mechanism that preserves communication performance while obfuscating sensitive spatial cues.

The privacy risks associated with channel measurements have been extensively demonstrated in the literature on device-free sensing and Wi-Fi-based activity recognition
\cite{abdelnasser2015wigest, wang2017device, jiang2021wifi, shenoy2022rf, zhu2025csi}. These works show that subtle CSI variations can reveal human presence, gestures, and even respiration patterns. However, most existing countermeasures, such as temporal smoothing, random antenna selection, or coarse quantization, offer only heuristic protection, lack formal privacy guarantees, and often degrade beamforming performance.

More recently, empirical studies have revealed that \emph{standard Wi-Fi beamforming feedback itself} can directly leak sensitive information. Liu \emph{et al.}~\cite{liu2024lendmeyourbeam} showed that the plaintext beamforming feedback specified in IEEE~802.11ac/ax, originally designed solely for link adaptationm exposes environment- and device-specific signatures that enable user identification, gesture recognition, and fine-grained localization. Their findings highlight a critical vulnerability: the right-singular subspace of the MIMO channel, when reported verbatim, effectively acts as a stable spatial fingerprint of the surrounding environment.
\subsection{Related Work}

Despite the rapid progress of wireless sensing and channel-aware inference, corresponding privacy and security safeguards have not kept pace. CSI and feature-level feedback streams on commercial off-the-shelf (COTS) Wi-Fi and cellular devices are now routinely exploited for activity recognition, localization, and environmental inference, yet commodity transceivers provide \emph{no built-in defenses} against such unintended inference or adversarial reconstruction. Existing mitigation approaches either rely on specialized hardware (e.g., reconfigurable surfaces or shielded arrays) or remain ad hoc, offering no analytical characterization of privacy leakage or performance loss.

A recent effort to obfuscate beamforming feedback via randomized angle perturbations~\cite{cominelli2024physical} represents an important step forward, but is limited to simple MIMO configurations and lacks a principled privacy interpretation. In particular, the interaction between link adaptation dynamics and injected randomness makes it difficult to isolate and quantify the resulting privacy gains. This motivates the key question we address:
\emph{Is it possible to design a software-level, standards-compliant CSI feedback mechanism that provides formal privacy guarantees while preserving the beamforming utility of existing Wi-Fi systems?}

\begin{figure}[t]
    \centering
    \includegraphics[width= 0.9\columnwidth]{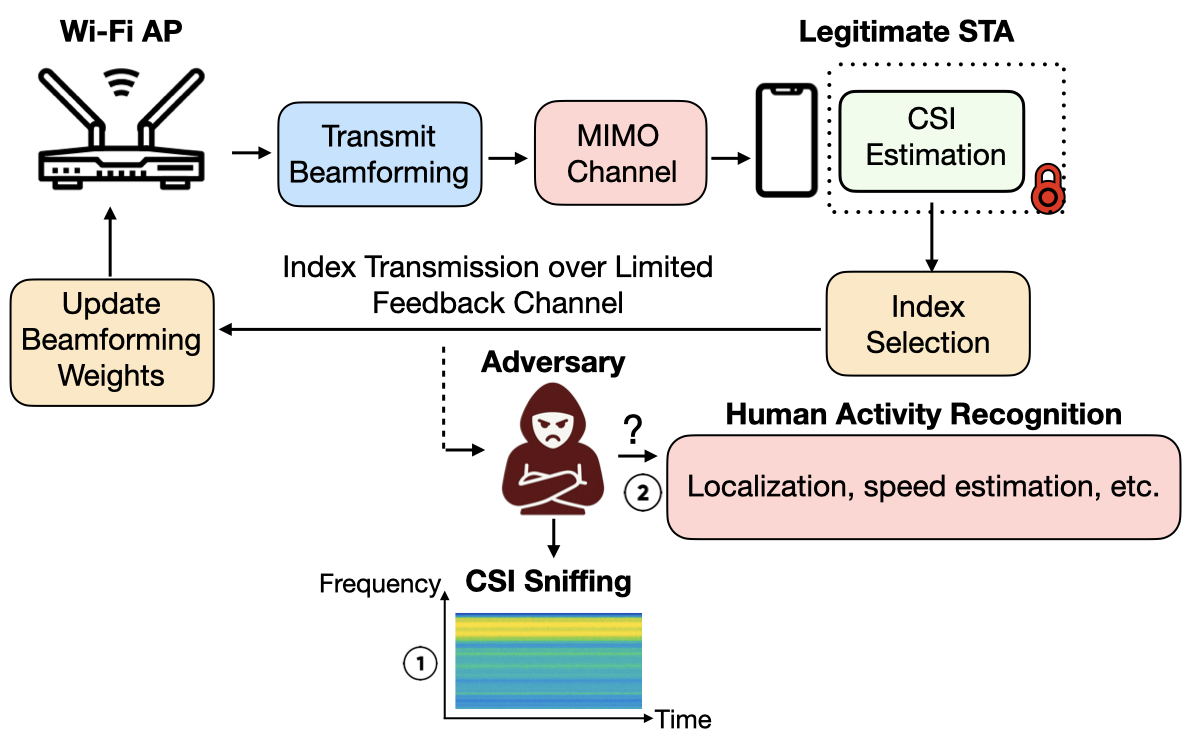}
    \caption{\small{An adversarial wireless setting where: $(1)$ a passive eavesdropper  leverages the physical-layer structure of received Wi-Fi signals—e.g., through spectrogram measurements or the rotation-angle parameters conveyed during CSI feedback, and $(2)$ infer sensitive user or environmental information without authorization.}}
    \label{fig:threat_model}
\end{figure}

\noindent \textbf{Summary of Contributions.}
In this work, we develop a {unified framework for privacy-preserving CSI feedback under differential privacy (DP) \cite{dwork2014algorithmic} } tailored to the Givens rotation and phase parameters used in IEEE~802.11  compressed beamforming feedback. Instead of modifying the CSI subspace via hidden rotations, we directly privatize the \emph{standard-reported} angular parameters by replacing deterministic quantization with an $\varepsilon$-DP stochastic quantization rule. The proposed mechanism is fully compatible with the 802.11 feedback structure and admits closed-form sensitivity characterizations for the angular mapping. To the best of our knowledge, this is the first work to
\begin{enumerate}
\item \emph{propose a DP-compliant CSI feedback scheme} that privatizes Givens and phase angles through carefully designed stochastic quantization while preserving standards compatibility;
\item \emph{derive analytical bounds} characterizing the degradation in beamforming utility induced by the DP perturbation of angular parameters;
\item \emph{conduct numerical simulations} demonstrating the privacy–utility trade-off, validating the analytical bounds, and showcasing the potential of the proposed scheme through adversarial speed-estimation attacks operating on privatized CSI reports.
\end{enumerate}

\noindent \textbf{Paper Organization.} The remainder of the paper is organized as follows. Section \ref{sec:system_model} introduces the system model and the privacy threat model. In Section \ref{sec:privacy_mechanism}, we propose our privacy-preserving stochastic quantization mechanism. Section \ref{sec:numerical_results} shows numerical results to confirm our findings.  Finally, Section \ref{sec:conclusion} concludes the paper and discuss future directions.

\begin{figure}[t]
\centering
\resizebox{\columnwidth}{!}{%
\begin{tikzpicture}[
    font=\scriptsize,
    >=latex,
    transform shape,
    node distance=4.5mm and 6.5mm,
    block/.style={draw, thick, rounded corners, align=center,
                  minimum width=1.55cm, minimum height=0.62cm, fill=gray!5},
    codebook/.style={draw, thick, dashed, rounded corners, align=center,
                     minimum width=1.55cm, minimum height=0.58cm, fill=gray!15}
]

\node[block] (tx) {AP (TX)\\ $N_t$ ants\\ $\widehat{\mathbf{V}}$};
\node[block, right=of tx] (H) {$\mathbf{H}$\\ Channel};
\node[block, right=of H] (rx) {STA (RX)\\ $N_r$ ants};
\node[block, right=of rx] (csi) {CSI proc.\\ Est.\,+\,SVD\\ $\Theta$};
\node[block, right=of csi] (dpq) {DP-SQ\\ $\widehat{\Theta}$};

\node[above=3.5mm of tx] (data) {$\mathbf{s}\in\mathbb{C}^{N_s}$};
\draw[->, thick] (data) -- (tx);

\draw[->, thick] (tx) -- (H);
\draw[->, thick] (H) -- (rx);
\draw[->, thick] (rx) -- (csi);
\draw[->, thick] (csi) -- (dpq);

\coordinate (fbmid) at ($(dpq.south)+(0,-6.5mm)$);
\draw[->, thick]
    (dpq.south) |- (fbmid) -|
    node[pos=0.52, below] {\scriptsize CSI feedback} (tx.south);

\node[codebook, below=6.5mm of csi] (cb) {Angles codebook};

\draw[dashed, <->, thick] (cb.north) -- (csi.south);
\draw[dashed, <->, thick] (cb.west) |- (tx.south);

\node[below=2.0mm of cb] {\scriptsize \textbf{Stochastic DP quantization in feedback}};

\end{tikzpicture}%
}
\caption{\small{Closed-loop CSI feedback architecture with $N_t$ transmit antennas, $N_r$ receive antennas, $N_s$ spatial streams, a shared angles codebook, and the proposed DP-SQ feedback quantization.}}
\label{fig:compact_csi_model}
\end{figure}

\section{System Model}
\label{sec:system_model}

In this section, we introduce the communication model, review the CSI feedback mechanism standardized for Wi-Fi beamforming, and present a concrete example illustrating how an adversary can exploit the reported feedback parameters to estimate the motion speed of a person moving within the Wi-Fi coverage area (e.g., inside a home or indoor environment).

\vspace{-0.05in}

\subsection{Communication Model}

As defined in the IEEE~802.11 standard, the establishment of a MIMO transmission in Wi-Fi proceeds in two primary stages: $(1)$ channel sounding, and $(2)$ precoder computation, described below. We consider a block-fading MIMO downlink channel
$\mathbf{H} \in \mathbb{C}^{N_r \times N_t}$ that remains constant over each coherence block. We assume for simplicity that the channel is narrowband. The access point (AP) is equipped with $N_t$ antennas and transmits $N_s$ spatial streams, while the station (STA) is equipped with $N_r$ receive antennas. The compact SVD of $\mathbf{H}$ is expressed as
\begin{align}
\mathbf{H} = \mathbf{U}\boldsymbol{\Sigma}\mathbf{V}^H.
\end{align}
During the sounding phase of a coherence block, the AP transmits a known
pilot matrix $\mathbf{S}\in\mathbb{C}^{N_t\times T_p}$ over $T_p$ pilot
symbols. Each pilot symbol is transmitted with power $P$, so that
$\frac{1}{T_p} \cdot \mathrm{tr}(\mathbf{S}\mathbf{S}^H) = P$. We adopt the standard orthogonal training design, i.e., 
$\mathbf{S}\mathbf{S}^H = P T_p \, \cdot \mathbf{I}_{N_t}$,
which requires $T_p \ge N_t$ and ensures optimal  least-squares (LS) estimation under
Gaussian noise. The STA receives
\begin{equation}
\mathbf{Y}
= \mathbf{H}\mathbf{S} + \mathbf{W},
\label{eq:pilot_received}
\end{equation}
where $\mathbf{W}\sim\mathcal{CN}(0, N_0 \mathbf{I}_{N_r T_p})$ is the receiver noise and $N_{0}$ is the noise power. The STA forms the LS estimate:
\begin{equation}
\widehat{\mathbf{H}}
= \mathbf{Y}\mathbf{S}^H (\mathbf{S}\mathbf{S}^H)^{-1}
= \mathbf{H} + \mathbf{W}\mathbf{S}^H (\mathbf{S}\mathbf{S}^H)^{-1}.
\label{eq:ls_estimate}
\end{equation}
Using the orthogonal training condition, the LS
estimate simplifies to
\begin{equation}
\label{eq:ls_simplified}
\widehat{\mathbf{H}}
= \frac{1}{P T_p}\, \cdot \mathbf{Y}\mathbf{S}^H,
\end{equation}
where the error covariance of this estimate is given as
\[
\mathbb{E}\!\left[
(\widehat{\mathbf{H}} - \mathbf{H})
(\widehat{\mathbf{H}} - \mathbf{H})^H
\right]
=
\frac{N_0}{P T_p}\, \cdot \mathbf{I}_{N_r},
\]
The estimate $\widehat{\mathbf{H}}$ is then used by the STA to generate compressed CSI feedback.

\subsection{Compressed CSI Feedback via Givens Angles}

Following the explicit beamforming feedback procedure, the STA computes the SVD of the estimated channel $\widehat{\mathbf{H}}$:
\begin{equation}
\widehat{\mathbf{H}}
= \widehat{\mathbf{U}}\,
\widehat{\boldsymbol{\Sigma}}\,
\widehat{\mathbf{V}}^{H},
\end{equation}
and extracts the right-singular subspace $\widehat{\mathbf{V}}
\in \mathbb{C}^{N_t \times N_s}$ responsible for downlink beamforming. To represent $\widehat{\mathbf{V}}$ efficiently, the standard uses a set of
angular parameters
\[
\Theta = \big\{ 
\phi_1,\dots,\phi_{N_t},
\; \psi_{i,j} :
1 \le i < j \le N_t
\big\},
\]
where $\phi_i \in [0,2\pi)$ are per-antenna phase angles, and $\psi_{i,j} \in [0,\tfrac{\pi}{2})$ are Givens rotation angles that 
describe the $(i,j)$ plane rotations. Each $\phi_i$ is quantized using $B_{\phi}$ bits and each $\psi_{i,j}$ using
$B_{\psi}$ bits, following the compressed beamforming report (CBR) format in
Annex N of IEEE~802.11ac/ax.  The resulting discrete set is
\begin{align}
\widehat{\Theta}
=
\big\{
\widehat{\phi}_1,\dots,\widehat{\phi}_{N_t},
\; \widehat{\psi}_{i,j}:
1 \le i < j \le N_t 
\big\} \nonumber 
\end{align}
is fed back to the AP. Using these quantized angles, the AP reconstructs the quantized precoder:
\begin{equation}
\widehat{\mathbf{V}}
=
\bigg(
\prod_{1 \le i < j \le N_t}
\mathbf{G}_{i,j}(\widehat{\psi}_{i,j})
\bigg)
\mathbf{D}(\widehat{\boldsymbol{\phi}})
\mathbf{E}_{N_s},
\label{eq:V_reconstruct}
\end{equation}
where $N_{s}$ is the number of transmit streams and 
\begin{align*}
\mathbf{G}_{i,j}(\psi_{i,j})
&= \mathbf{I}, \text{except for $(i,j)$th block}
\begin{bmatrix}
\cos\psi_{i,j} & -\sin\psi_{i,j} \\
\sin\psi_{i,j} & \cos\psi_{i,j}
\end{bmatrix}, \\
\mathbf{D}(\boldsymbol{\phi})
&= 
\mathrm{diag}
\left(
e^{j\phi_1},\dots,e^{j\phi_{N_t}}
\right), \\
\mathbf{E}_{N_s}
&=
\begin{bmatrix}
\mathbf{I}_{N_s} \\
\mathbf{0}
\end{bmatrix}
\in \mathbb{C}^{N_t \times N_s}.
\end{align*}

In the special case $N_t = 2$ and $N_r = 1$, the CSI feedback reduces to a 
single Givens rotation angle $\psi_{1,2}$. The associated Givens matrix is
\[
\mathbf{G}_{1,2}(\psi_{1,2})
=
\begin{bmatrix}
\cos\psi_{1,2} & -\sin\psi_{1,2} \\
\sin\psi_{1,2} & \cos\psi_{1,2}
\end{bmatrix}.
\]
Using the quantized angles $(\widehat{\psi}_{1,2},\,\widehat{\phi}_1,\,\widehat{\phi}_2)$, the AP reconstructs:
\[
\widehat{\mathbf{V}}
=
\mathbf{G}_{1,2}(\widehat{\psi}_{1,2})
\mathbf{D}(\widehat{\boldsymbol{\phi}})
\mathbf{E}_1
=
\begin{bmatrix}
\cos(\widehat{\psi}_{1,2})\,e^{j\widehat{\phi}_1} \\
\sin(\widehat{\psi}_{1,2})\,e^{j\widehat{\phi}_2}
\end{bmatrix}.
\]
This $\widehat{\mathbf{V}} \in \mathbb{C}^{2\times 1}$ is then used as the
downlink beamforming vector for the single transmitted stream.

\subsection{Privacy Threat Model}
\label{subsec:privacy_threat_model}

We next consider a passive wireless adversary (see Fig.~\ref{fig:threat_model}) that does not interfere with the
protocol, but continuously \emph{observes} explicit CSI feedback exchanged
between a STA and an AP. The adversary is assumed
to:
(i) know the IEEE~802.11ac/ax compressed beamforming format;
(ii) be able to parse the reported Givens rotation and phase angles from
each feedback report; and
(iii) have sufficient computational resources to perform offline signal
processing and learning on the collected CSI logs. The adversary's goal is not to disrupt communication, but to infer
\emph{sensitive side information} about the environment and users from the
temporal evolution of the reported CSI. In this work, we focus on
\emph{activity and motion speed inference}: by tracking how the reported
Givens angles evolve over time, the adversary attempts to estimate whether
a user is stationary, performing small gestures (e.g., typing), or moving
at higher speeds (e.g., walking), and to recover coarse speed information.
Crucially, this attack operates \emph{purely} on the plaintext CSI feedback
already present in 802.11-compliant systems and requires no additional
hardware beyond a receiver capable of decoding control frames. Moreover, because the CSI is already embedded in the reported feedback, the attacker can bypass complex processing and does not need to directly observer the STA-AP channel (i.e., where typically the attacker would need to first estimate their own channel to the STA and then infer the AP–STA channel).  \\

\noindent \textbf{An Example for Adversarial Speed Estimation.} To illustrate the privacy risk, consider the simplest explicit-beamforming
configuration with $N_t=2$ transmit antennas and $N_r=1$ receive antenna (Table.~\ref{tab:chan_params} provides more details on the experimental setup).  In this case the right-singular vector of the channel is parameterized by a single
Givens rotation angle $\psi_{1,2}$ and two per-antenna phases.  The AP uses
the quantized angle $\widehat{\psi}_{1,2}(t)$ reported by the STA, but the
same value is also observable to a passive eavesdropper.  A passive adversary collects a time series of Givens angles
\(
\{\widehat{\psi}_{1,2}(t_k)\}_{k=1}^M
\)
from successive CSI feedback packets.  
Mirroring classical micro-Doppler processing used in CSI-based sensing schemes (e.g.,~\cite{cominelli2024physical}), we now describe how an observer can convert a time series of CSI reports into Doppler and speed estimates. We consider a  block-fading MIMO channel with $N_{\mathrm{sc}}$ active subcarriers and $N_{\mathrm{snap}}$ CSI snapshots. At snapshot index $n \in \{0,\dots,N_{\mathrm{snap}}-1\}$, the discrete-time downlink channel is denoted as $\mathbf{H}[n]$, and the corresponding CSI report is generated at time $t_n = n T_{\mathrm{CSI}}$, where $T_{\mathrm{CSI}}$ is the CSI reporting interval. In practice, both the access point and a passive eavesdropper operate on an \emph{effective} CSI obtained by projecting $\mathbf{H}[n]$ onto fixed transmit/receive beams and, in the explicit-beamforming case,  reconstructing those beams from compressed CSI feedback. We denote the resulting frequency-domain CSI per subcarrier by
\begin{equation}
  h[k,n] \in \mathbb{C}, 
  \quad k = 1,\dots,N_{\mathrm{sc}},
  \label{eq:scalar_csi_def}
\end{equation}
where $k$ denotes subcarrier index.

To obtain a single phase trajectory that captures the dominant Doppler over time, the observer first aggregates the CSI,
\begin{equation}
  \widetilde{h}[n]
  =
  \sum_{k=1}^{N_{\mathrm{sc}}} w_k\, h[k,n],
  \quad
  \sum_{k=1}^{N_{\mathrm{sc}}} w_k = 1,
  \label{eq:wideband_surrogate}
\end{equation}
where $w_k$'s are fixed combining non-negative weights (e.g., proportional to the average subcarrier SNR). This yields a single complex-valued sample $\widetilde{h}[n]$ per CSI snapshot. The instantaneous phase of this sample is
\begin{equation}
  \phi[n] = \arg\bigl(\widetilde{h}[n]\bigr), \nonumber 
\end{equation}
which is then \emph{unwrapped} across $n$ to remove $2\pi$ discontinuities, producing a smooth phase trajectory $\widetilde{\phi}[n]$ as a function of time $t_n$. When the narrowband channel is dominated by an effective Doppler shift $f_D$, the sample can be approximated as
\begin{equation}
  \widetilde{h}[n]
  \approx
  A \exp \bigl(j 2\pi f_D t_n\bigr), \nonumber
\end{equation}
for some complex amplitude $A$. Here, the unwrapped phase evolves approximately linearly in time as follows
\begin{equation}
  \widetilde{\phi}[n]
  \approx 2\pi f_D t_n + \phi_0,
\end{equation}
with $\phi_0$ a constant phase offset. The observer (access point or eavesdropper) estimates the slope by fitting a straight line to the phase–time pairs $\{(t_n,\widetilde{\phi}[n])\}$ via least squares:
\begin{equation}
  \widetilde{\phi}[n] \approx a t_n + b, \nonumber 
\end{equation}
where $a$ and $b$ are obtained from a standard linear regression. Identifying $a \approx 2\pi f_D$, the Doppler estimate is
\begin{equation}
  \widehat{f}_D = \frac{a}{2\pi}.
  \label{eq:doppler_estimate}
\end{equation}

Next, let $f_c$ denote the carrier frequency and $\lambda = c / f_c$ to be the corresponding wavelength, where $c$ is the speed of light. Because we observe the \emph{one-way} propagation channel (as opposed to a two-way radar echo), the effective radial speed is related to the Doppler shift by
\begin{equation}
  \widehat{v} = \lambda \, \widehat{f}_D.
  \label{eq:speed_estimate}
\end{equation}
Thus yielding a scalar speed estimate from each CSI sequence. To track speed variations over time and obtain a time-varying speed trajectory $\widehat{v}(t)$, the same phase-slope estimator can be applied in a short-time (sliding-window) fashion. Specifically, for each center index $m$ the observer considers a local window $\mathcal{W}_m = \{n : |n-m| \leq W/2\}$ of $W$ CSI snapshots, forms the complex sample $\widetilde{h}[n]$ as in Eqn.~\eqref{eq:wideband_surrogate} for all $n \in \mathcal{W}_m$, unwraps the phase
\[
  \widetilde{\phi}[n] = \arg\bigl(\widetilde{h}[n]\bigr),
  \quad n \in \mathcal{W}_m,
\]
and fits a local line
\[
  \widetilde{\phi}[n] \approx a_m t_n + b_m, \quad n \in \mathcal{W}_m,
\]
via least squares. The local Doppler and speed estimates at time $t_m$ are then
\begin{equation}
  \widehat{f}_D[m] = \frac{a_m}{2\pi},
  \quad
  \widehat{v}[m]   = \lambda\, \widehat{f}_D[m].
\end{equation}

\begin{figure}[t]
\centering
\includegraphics[width=\columnwidth]{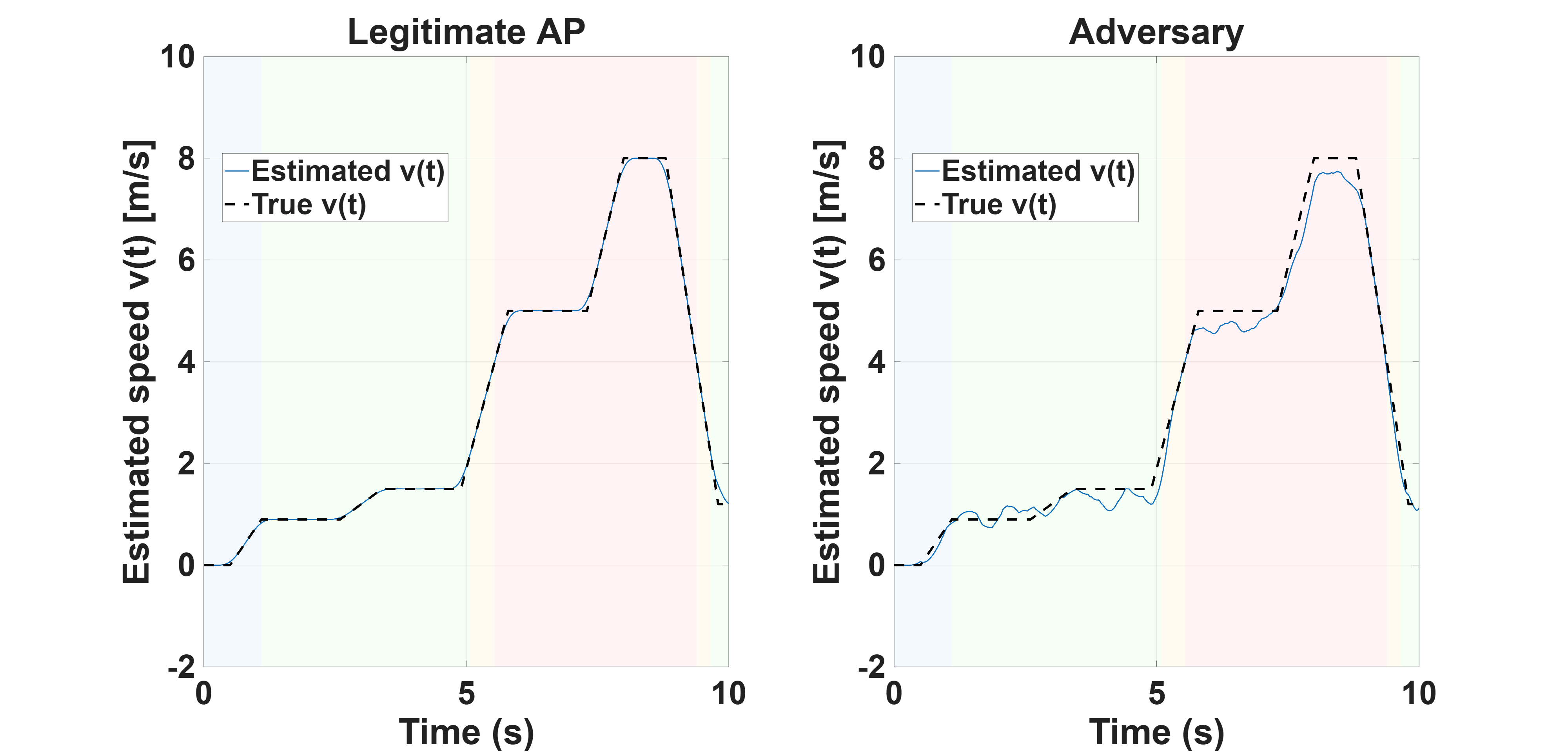}
\centering
    \caption{\small{Information leakage from the compressed CSI packets. The adversary is assumed to have a lower SNR then the legitimate AP ($20$ dB vs $5$ dB) but is still able to accurately reconstruct the full CSI and estimate the STA's speed from the compressed feedback packets.}}
    \label{fig:threat_example}
\end{figure}

Crucially, this estimation pipeline is agnostic to how the effective CSI $h[k,n]$ is obtained: it can be computed from raw downlink pilots, from beamformed CSI at the receiver, or reconstructed from compressed CSI feedback (e.g., Givens-parameter reports) using the standard IEEE~802.11 beamforming structure. As a result, by continuously tracking the complex CSI associated with the reported beams, a passive eavesdropper can infer the user's motion speed over time without requiring explicit access the access point's or STA information as shown Fig.~\ref{fig:threat_example} illustrates this attack. Because our approach operates exclusively on the standard Givens rotation and phase angles already utilized in IEEE 802.11 explicit feedback, it requires no hardware modifications or alterations to existing channel sounding procedures. Consequently, it serves as a seamless, drop-in privacy enhancement for commercial transceivers that rigorously preserves the required orthonormality and structure of the beamforming matrix.\\

\begin{table}[t]
\centering
\caption{\small{Simulation parameters for the synthetic CSI generation.}}
\label{tab:chan_params}
\begin{tabular}{l l l}
\toprule
\textbf{Field} & \textbf{Value} & \textbf{Description} \\
\midrule
\verb|NumTx|            & 2     & $\#$ of AP transmit antennas ($N_t$) \\
\verb|NumRx|            & 1     & $\#$ of STA receive antennas ($N_r$) \\
\verb|ChannelBandwidth| & CBW20 & 20 MHz bandwidth \\
\verb|NumSTS|           & 1     & Spatial streams ($N_{s}$) \\
\verb|NumPackets|       & 5000  & CSI snapshots \\
\verb|NumPaths|         & 10     & Multipath components \\
\verb|MaxDelaySamples|  & 20    & Max tap delay \\
\verb|CenterFreqHz|     & $5.785{\times}10^{9}$ & Carrier freq.\ ($f_c$) \\
\verb|IntervalSec|      & $10^{-3}$ sec           & CSI Measurement Interval ($\Delta t$) \\
\verb|KFactor_dB|       & 4 dB & Rician $K$-factor \\
\verb|VelocityAngleRad| & 0 rad & User direction \\
\bottomrule
\end{tabular}
\end{table}

\begin{figure}[t]
\centering
\includegraphics[width= \columnwidth]{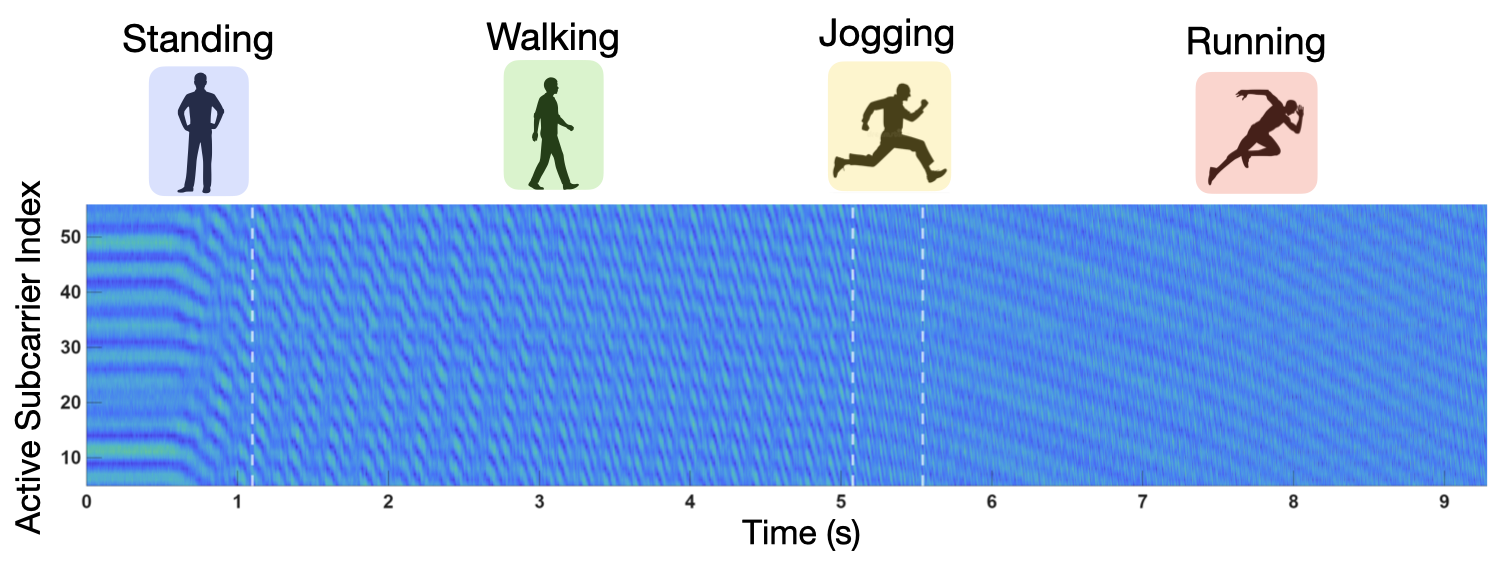}
\centering
    \caption{\small{An example for CSI amplitude spectrogram illustrating activity-dependent channel dynamics for a user moving within an indoor environment. Distinct Doppler patterns correspond to transitions between standing/slow motion, walking, jogging, and running, reflecting how user activity modulates the wireless channel over time.}}
    \label{fig:spectrogram_example}
\end{figure}

\vspace{-0.1in}

\section{Stochastic Quantization for Privacy-Preserving Beamforming}
\label{sec:privacy_mechanism}

In this section, we describe the proposed privacy-preserving mechanism for the 
beamforming transmit matrix $\mathbf{V}$.  
The key idea is to apply a randomized quantization procedure \footnote{ We want to shed the light that while the receiver noise inherently introduces randomness into the feedback process, 
it constitutes an {uncontrolled} and {unpredictable} source of privacy. 
Although such noise may provide incidental obfuscation of the beamforming feedback, 
its distribution and temporal correlation are determined by the physical channel conditions and hardware impairments, 
and therefore cannot be relied upon to provide formal or quantifiable privacy guarantees.
} to the 
Givens–rotation angles that parametrize $\mathbf{V}$ in the IEEE~802.11-style 
compressed beamforming feedback.  
Stochastic quantization introduces controlled randomness into each angle, which 
both masks fine-grained channel structure and enables formal DP
guarantees when the quantization probabilities are appropriately biased.  
We first review the basic stochastic quantization model and its MSE properties, which will serve as the building block for our proposed DP mechanism.

\subsection{Stochastic Quantization}

Consider a uniform scalar quantizer over the interval $ \bar{a} \triangleq  [a_{\min},a_{\max}]$ with 
$L = 2^{B}$ reconstruction levels, where $B$ denotes the number of quantization bits.  Let $\mathcal{Q} $ be the set of the quantization levels, i.e., $\mathcal{Q} = \{q_1,q_2,\ldots,q_L\}$, where each level $ q_j$ is expressed as
\begin{align} 
\qquad q_j = a_{\min} +   \frac{j-1}{L-1} \cdot \bar{a}, \quad j \in \{1,\dots,L-1\},
\nonumber 
\end{align}
where $q_{1} < q_{2} < \cdots < q_{L}$. For any input $a $, let $q_i$ and $q_{i+1}$ denote the two nearest 
quantization levels such that
\[
q_i \le a \le q_{i+1}, \quad i \in \{1,\dots,L-1\}.
\]
A \emph{stochastic} quantizer maps $a$ to one of these two levels according to the following probabilstic mapping:
\[
Q_b(a)=
\begin{cases}
q_i,     & \text{w.p.} \quad p,\\[0.7ex]
q_{i+1}, & \text{w.p.} \quad 1-p,
\end{cases}
\]
where $p\in (0,1]$ is a design parameter. Let the quantization step be $\Delta = q_{i+1}-q_i$.  
If $a$ lies uniformly within the quantization cell (i.e., \ $a = q_i + r$ with 
$r\sim\mathrm{Unif}[0,\Delta]$), then the MSE can be readily shown to be
\begin{align}
\mathbb{E}\!\left[(Q_b(a)-a)^{2}\right]
= \frac{\Delta^{2}}{12}\, \cdot \bigl(4 - 6p + 6p^{2}\bigr).
\label{eqn:MSE_determinstic}
\end{align}

We next show how to design the stochastic quantization parameter 
$p$ to ensure provable privacy guarantees. Let us first formalize differential privacy in this setting as follows.

\subsection{Differential Privacy}

We consider an adversary whose observable is the standard feedback payload\footnotetext{
While channel sounding may leak some activity-related information, analog CSI is known to attenuate significantly with wall penetration and distance, whereas digital beamforming feedback remains robust and can be passively observed via packet sniffing.
}: namely, the vector of quantized Givens parameters (phase and rotation angles, or equivalently their codeword indices) derived from the estimated channel. As illustrated in Subsection~\ref{subsec:privacy_threat_model}, an eavesdropper can exploit these reported quantities to infer user activity or motion patterns from the underlying channel realizations.

We next recall the standard notion of differential privacy (DP), which provides a formal measure of indistinguishability between outputs corresponding to neighboring inputs.

\begin{definition}[(\(\varepsilon,\delta\))-DP \cite{dwork2014algorithmic}]
A randomized mechanism $\mathcal{M}$ is said to be $(\varepsilon,\delta)$-differentially private if for any two inputs $\mathbf{V},\mathbf{V}'$ and any measurable set $E$ in the output space,
\begin{align}
\Pr[\mathcal{M}(\mathbf{V})\in E]
\le
e^{\varepsilon}\Pr[\mathcal{M}(\mathbf{V}')\in E]+\delta,
\end{align}
where $\varepsilon>0$ and $\delta\ge 0$ are privacy parameters. Smaller values of $(\varepsilon,\delta)$ correspond to stronger privacy guarantees. The special case $\delta=0$ is referred to as pure $\varepsilon$-DP.
\label{def:DP_beamforming}
\end{definition}

In the context of beamforming feedback, directly enforcing global $(\varepsilon,\delta)$-DP over the continuous angular domain would require full-support randomization across all reconstruction levels, which can significantly degrade the utility of the reported beamforming parameters. Instead, we adopt a localized privacy notion aligned with the quantized representation of the feedback, as detailed next.

\subsection{DP Stochastic Quantization (DP-SQ) Mechanism}

Consider a true Givens angle $a$ to be quantized on a uniform grid $\mathcal{Q}$ with spacing $\Delta$. For the phase and mixing angles, we have
\[
\phi\in[-\pi,\pi),\quad \Delta_\phi=\tfrac{2\pi}{2^{B_\phi}},
\qquad
\psi\in[0,\tfrac{\pi}{2}],\quad \Delta_\psi=\tfrac{\pi/2}{2^{B_\psi}}.
\]
Let $(q_i,q_{i+1})$ denote the two nearest reconstruction levels to $a$, i.e.,
\[
q_i \le a \le q_{i+1}, \quad i\in\{1,\dots,|\mathcal{Q}|-1\},
\]
where distances are interpreted in the circular sense for $\phi$ and in the standard (clamped) sense for $\psi$.

We define a local adjacency relation on the angular domain: two angles $a$ and $a'$ are said to be adjacent, denoted $a\sim a'$, if they belong to the same quantization interval, $a,a'\in[q_i,q_{i+1}] \quad \text{for some } i$.

\noindent \textbf{DP-SQ mechanism.}
The proposed stochastic quantization mechanism releases a randomized reconstruction $\widehat{a}\in\{q_i,q_{i+1}\}$ according to
\begin{equation}
\widehat{a}=
\begin{cases}
q_i, &\text{w.p.} \quad 
p^\star(\varepsilon_a)=\dfrac{e^{\varepsilon_a}}{e^{\varepsilon_a}+1},\\[0.8ex]
q_{i+1}, &\text{w.p.} \quad 1-p^\star(\varepsilon_a),
\end{cases}
\label{eq:dpsq-rule}
\end{equation}
where $\varepsilon_a$ is a design parameter controlling the level of randomization (e.g., $\varepsilon_\phi$ for the phase and $\varepsilon_\psi$ for the mixing angle).

The above mechanism preserves the standard two-level stochastic quantization structure while introducing controlled randomization between adjacent reconstruction levels. In particular, the released value $\widehat{a}$ depends only on the local quantization cell $[q_i,q_{i+1}]$ and not on the exact location of $a$ within the cell.

\noindent \textbf{Privacy guarantee.} Under the cell-wise adjacency relation, the DP-SQ mechanism provides the following privacy guarantee.

The above guarantee implies that all angles within the same quantization interval are statistically indistinguishable, thereby limiting the resolution at which the underlying angle can be inferred. The parameter $\varepsilon_a$ governs a local privacy--utility tradeoff by controlling the bias between the two adjacent reconstruction levels: as $\varepsilon_a \to 0$, the mechanism approaches uniform randomization, while as $\varepsilon_a \to \infty$, it approaches deterministic quantization. We emphasize that this privacy notion is local to each quantization cell. Achieving global $\varepsilon$-local DP over arbitrary input pairs would require full-support randomization across all reconstruction levels (e.g., via $M$-ary randomized response), which would significantly increase angular distortion and degrade beamforming performance. The proposed DP-SQ mechanism instead focuses on \emph{localized} randomization that preserves the structure and efficiency of standard CSI feedback representations.

\noindent \textbf{Utility.} We next introduce an intermediate lemma to quantify the distortion introduced by the DP-SQ mechanism.

\begin{lemma}
Let each quantizer have step size
$\Delta_\phi = \tfrac{2\pi}{2^{B_\phi}}$ and
$\Delta_\psi = \tfrac{\pi/2}{2^{B_\psi}}$.
Now, define the term
$\kappa(\varepsilon) \triangleq  \frac{e^{\varepsilon}-1}{e^{\varepsilon}+1}$.
The proposed privacy mechanism yields the following MSE:
\begin{align}
\label{eq:angle-mse-exact}
\mathbb{E}\!\big[(\widehat{a}-a)^2\big]
= \frac{\Delta^2}{12}\! \cdot \left(4 - 3\,\kappa(\varepsilon)\right).
\end{align}
\end{lemma}
\noindent It is worth highlighting that the above expression recovers the MSE guarantee of the deterministic quantization when there is no privacy constraints, i.e., $\varepsilon \rightarrow \infty$.

\begin{proof}
Consider a uniform mid-rise quantizer with step size $\Delta$. For a given
true angle $a$, let $r$ denote its distance to the nearest quantization
level. Then
$r := \operatorname{dist}(a,\text{nearest level}) \in [0,\Delta/2]$.
Under the standard high–resolution assumption, $a$ is uniformly distributed
within its quantization cell, so $r \sim \mathrm{Unif}\bigl([0,\Delta/2]\bigr)$. With privacy parameter $\varepsilon>0$, the DP stochastic quantizer outputs the nearest level with probability $p^\star(\varepsilon)$, and the {farther} neighbor (at distance $\Delta-r$) with probability
$1-p^\star(\varepsilon)$. Conditioned on $r$, the MSE is
\[
\mathbb{E}\big[(\widehat a-a)^2 \mid r\big]
= p^\star(\varepsilon)\,r^2
+ \bigl(1-p^\star(\varepsilon)\bigr)(\Delta-r)^2.
\]
The pdf of $r$ is $f_r(x)=\tfrac{2}{\Delta}$ on $[0,\Delta/2]$, hence
\begin{align*}
\mathbb{E}[r]
&= \int_0^{\Delta/2} x\, \cdot \frac{2}{\Delta}\,dx
= \frac{2}{\Delta}\cdot \frac{x^2}{2}\Big|_0^{\Delta/2}
= \frac{\Delta}{4},\\[0.5ex]
\mathbb{E}[r^2]
&= \int_0^{\Delta/2} x^2\, \cdot \frac{2}{\Delta}\,dx
= \frac{2}{\Delta}\cdot \frac{x^3}{3}\Big|_0^{\Delta/2}
= \frac{\Delta^2}{12}.
\end{align*}
Similarly,
\begin{align*}
\mathbb{E}\big[(\Delta-r)^2\big]
&= \Delta^2 - 2\Delta\,\mathbb{E}[r] + \mathbb{E}[r^2]\\
&= \Delta^2 - \frac{\Delta^2}{2} + \frac{\Delta^2}{12}
= \frac{7\Delta^2}{12}.
\end{align*}
Now average $\mathbb{E}[(\widehat a-a)^2\mid r]$ over $r$, we get
\begin{align*}
\mathbb{E}\big[(\widehat a-a)^2\big]
&= p^\star(\varepsilon)\,\mathbb{E}[r^2]
+ \bigl(1-p^\star(\varepsilon)\bigr)\,\mathbb{E}\big[(\Delta-r)^2\big]\\[0.5ex]
&= \frac{\Delta^2}{12}\, \cdot \bigl(7 - 6\,p^\star(\varepsilon)\bigr).
\end{align*}
Substituting $p^\star(\varepsilon)=\tfrac{1+\kappa(\varepsilon)}{2}$ gives
\begin{align*}
7 - 6\,p^\star(\varepsilon)
&= 7 - 3\bigl(1+\kappa(\varepsilon)\bigr)
= 4 - 3\,\kappa(\varepsilon),
\end{align*}
so
\[
\mathbb{E}\big[(\widehat a-a)^2\big]
= \frac{\Delta^2}{12}\, \cdot \bigl(4 - 3\,\kappa(\varepsilon)\bigr),
\]
which is the claimed expression.
\end{proof}

\begin{figure*}[t]
    \centering
    \includegraphics[width= 0.7\textwidth]{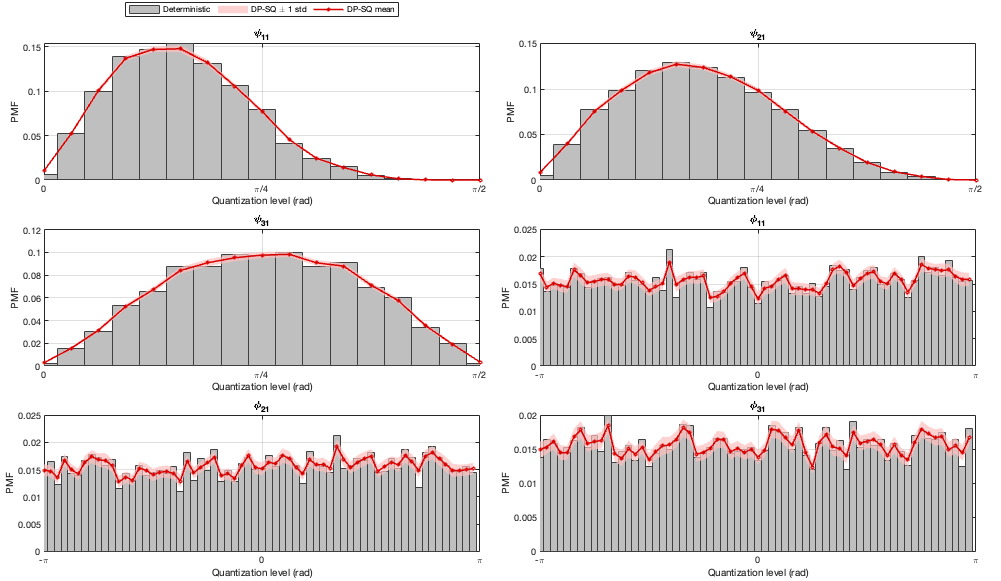}
    \caption{\small{Empirical PMFs of the quantized Givens angles under deterministic quantization and the proposed DP-SQ mechanism for single-stream beamforming with $(N_t,N_r)=(4,2)$, $(B_\phi,B_\psi)=(6,4)$, privacy parameters $(\varepsilon_\phi,\varepsilon_\psi)=(0.8,0.8)$, and $T=100$ independent DP-SQ realizations. Each subplot corresponds to one reported angle in the compressed feedback parameterization. The gray bars denote the deterministic PMF over reconstruction levels, while the red curves represent the mean PMF produced by DP-SQ across repeated runs, and the shaded regions indicate one standard deviation. The underlying raw angle samples are fixed across runs, and only the DP-SQ mechanism is randomized, thereby isolating the variability due solely to privacy-induced randomness.}}
    \label{fig:angle_pmf_variation}
\end{figure*}

Fig.~\ref{fig:angle_pmf_variation} illustrates the variability induced by the proposed DP-SQ mechanism relative to deterministic quantization. In contrast to the deterministic scheme, which yields a fixed PMF over reconstruction levels, DP-SQ introduces stochasticity through randomized selection between adjacent quantization points.  The resulting mean PMF and its associated standard deviation show that DP-SQ preserves the dominant structure of the deterministic angle distribution while causing only a localized spreading of probability mass around nearby reconstruction levels.

We next quantify how the DP-SQ scheme of
Givens angles perturbs the $N_s$-dimensional beamforming subspace.
Let $\mathbf V^\star\!\in\!\mathbb C^{N_t\times N_s}$ denote the ideal
right-singular matrix and
$\mathbf P^\star=\mathbf V^\star(\mathbf V^\star)^H$ its associated projector.
The quantized precoder generated from DP-SQ angles is denoted
$\widehat{\mathbf V}$ with projector $\widehat{\mathbf P}$. With the DP quantization mechanism established, we now formalize its impact on beamforming performance. The following theorem characterizes the transmit-side subspace distortion induced by our privacy-preserving angular quantization scheme.

\vspace{0.1cm}
\begin{theorem}[Expected Subspace Distortion under DP-SQ]
\label{thm:dpsq-utility-analytic-general-Ns}
Let $\mathbf V^\star=[\mathbf v_1^\star,\dots,\mathbf v_{N_s}^\star]$ have
orthonormal columns and define the subspace
$\mathcal S^\star=\operatorname{span}\{\mathbf V^\star\}$ with projector
$\mathbf P^\star$.  
Under the standard Givens parametrization of $\mathbf V^\star$, column $i$
depends on
\[
N_\psi(i)=N_\phi(i)=N_t-i,
\quad i=1,\dots,N_s,
\]
mixing angles $\{\psi_{i,k}\}$ and phase angles $\{\phi_{i,k}\}$.  
Thus the total number of mixing and phase angles is
$N_{\rm tot}
= \sum_{i=1}^{N_s}(N_t-i)
= N_sN_t - \frac{1}{2}{N_s(N_s+1)}$. Let $\bar{\boldsymbol\theta}$ denote the nearest–bin (deterministically)
quantized angles, and define
\[
\mathbf V_{\rm q}=\mathbf V(\bar{\boldsymbol\theta}),
\quad
\mathbf P_{\rm q}=\mathbf V_{\rm q}\mathbf V_{\rm q}^H.
\]
The deterministic quantization floor is defined as
\[
d_{\rm q}^2
\triangleq d_{\rm chord}^2(\mathcal S^\star,\mathcal S_{\rm q})
=\frac{1}{2} \cdot \|\mathbf P^\star-\mathbf P_{\rm q}\|_F^2.
\]

Now apply our \emph{independent} DP-SQ mechanism to each angle, yielding
(privatized) stochastic angles $\widehat{\boldsymbol\theta}$ and precoder
$\widehat{\mathbf V}=\mathbf V(\widehat{\boldsymbol\theta})$ with projector
$\widehat{\mathbf P}$, where each mixing and phase angle satisfies:
\[
\sigma_\phi^2
= \frac{\Delta_\phi^2}{12} \cdot \bigl(4-3\,\kappa(\varepsilon_\phi)\bigr),
\sigma_\psi^2 
= \frac{\Delta_\psi^2}{12} \cdot \bigl(4-3\,\kappa(\varepsilon_\psi)\bigr),
\]
with step sizes $\Delta_\phi,\Delta_\psi$ and DP bias factor
$\kappa(\varepsilon)
= \frac{e^\varepsilon - 1}{e^\varepsilon + 1}$.

Then the expected squared chordal distance between original subspace $\mathcal S^\star$ and the
 perturbed subspace $\widehat{\mathcal S}$  via DP-SQ  mechanism satisfies:
\begin{align}
\mathbb E\!\left[
d_{\rm chord}^2(\mathcal S^\star,\widehat{\mathcal S})
\right]
&\le
d_{\rm q}^2
+
2  N_s\,  N_{\rm tot} 
\big(\sigma_\psi^2+\sigma_\phi^2\big) \triangleq D(\epsilon_\psi,\epsilon_\phi).
\label{eq:dpsq-chordal-global-Ns-tight-self} 
\end{align}
\end{theorem}

\noindent \textbf{Proof Sketch.} For any two $N_s$-dimensional subspaces,
$d_{\rm chord}^2(\mathcal S_1,\mathcal S_2)
=\tfrac12\|\mathbf P_1-\mathbf P_2\|_F^2$.
Using the triangle inequality,
\[
d_{\rm chord}^2(\mathcal S^\star,\widehat{\mathcal S})
\le d_{\rm q}^2+\tfrac12\|\mathbf P_{\rm q}-\widehat{\mathbf P}\|_F^2.
\]
A standard identity gives 
$\|\mathbf P_{\rm q}-\widehat{\mathbf P}\|_F\le2\|\mathbf V_{\rm q}-\widehat{\mathbf V}\|_F$, 
so
\[
\tfrac12\|\mathbf P_{\rm q}-\widehat{\mathbf P}\|_F^2
\le 2\sum_{i=1}^{N_s}\|\mathbf v_{i,\rm q}-\widehat{\mathbf v}_i\|_2^2.
\]
For unit vectors, $\|\mathbf a-\mathbf b\|_2^2\le2\sin^2\angle(\mathbf a,\mathbf b)$.
Let $d_i$ be the one dimensional chordal distortion of column~$i$. Then
\[
\mathbb E[d_{\rm chord}^2]\le d_{\rm q}^2+4\sum_{i=1}^{N_s}\mathbb E[d_i^2].
\]
Each column $i$ is generated by $N_\psi(i)$ mixing angles and $N_\phi(i)$ phases.
Let $\mathbf v_i^{(m)}$ denote the vector after perturbing $m$ angles.
At each stage only a $2\times2$ block changes, we can further show that
\[
\|\mathbf v_i^{(m)}-\mathbf v_i^{(m-1)}\|_2^2
\le
8\!  \left(
\sin^2\tfrac{\Delta\psi_{i,m}}{2}
+\sin^2\!\Psi_{i,+}\,\sin^2\tfrac{\Delta\phi_{i,m}}{2}
\right).
\]
Summing stages and applying Cauchy--Schwarz,
\[
d_i^2\le2\sum_m\|\mathbf v_i^{(m)}-\mathbf v_i^{(m-1)}\|_2^2.
\]
Using the fact $\sin^2x\le x^2$ and the DP-SQ per‐angle MSEs,
$\mathbb E[\Delta\psi_{i,m}^2]=\sigma_\psi^2$, 
$\mathbb E[\Delta\phi_{i,m}^2]=\sigma_\phi^2$, we obtain
\[
\mathbb E[d_i^2]
\le
2 \cdot N_\psi(i) \cdot \sigma_\psi^2
+2\, \cdot \mathbb E[\sin^2\!\Psi_{i,+}]\, \cdot N_\phi(i) \cdot \sigma_\phi^2.
\]
Substituting back gives
\[
\mathbb E[d_{\rm chord}^2]
\le
d_{\rm q}^2
+4\sum_{i=1}^{N_s}\Big(
N_\psi(i) \cdot \sigma_\psi^2
+\mathbb E[\sin^2\! \Psi_{i,+}]\,\cdot N_\phi(i) \cdot \sigma_\phi^2
\Big).
\]
Under the Annex--N parametrization, 
$N_\psi(i)=N_\phi(i)=N_t-i$, so the total number of angles is 
$N_{\rm tot}=N_sN_t-\tfrac12N_s(N_s+1)$.
Define $\overline w_\phi=\frac1{N_s}\sum_{i=1}^{N_s}\mathbb E[\sin^2\!\Psi_{i,+}]\le1$.
Then
\[
\mathbb E[d_{\rm chord}^2]
\le
d_{\rm q}^2
+2 \cdot N_s \cdot N_{\rm tot} \cdot \big(\sigma_\psi^2+\overline w_\phi\, \cdot \sigma_\phi^2\big),
\]
which establishes the proof of the theorem.  \\

\begin{remark}
The quantities $\sigma_\psi^2$ and $\sigma_\phi^2$ correspond to the per-angle mean squared errors induced by the DP stochastic quantization mechanism, thereby explicitly linking the privacy parameter $\epsilon$ to the resulting beamforming distortion.
\end{remark}

We next show that the proposed DP-SQ mechanism preserves the orthonormality structure of the SVD-based beamforming matrix.

\noindent\textbf{Orthonormality under DP-SQ Mechanism.}
To make the argument concrete, consider the single-stream case with $N_t=2$
and $N_r=1$.
In this setting, any unit-norm beamforming vector admits the angular
parametrization
\begin{equation}
\mathbf v(\psi,\phi)
=
\begin{bmatrix}
\cos\psi \\
e^{j\phi}\sin\psi
\end{bmatrix},
\qquad
\psi\in[0,\tfrac{\pi}{2}],\ \phi\in[-\pi,\pi).
\end{equation}
Under the proposed DP-SQ mechanism, the angles $(\psi,\phi)$ are stochastically
quantized to neighboring grid points
$(\widehat{\psi},\widehat{\phi})\in\mathcal Q$ according to
\eqref{eq:dpsq-rule}, and the transmitted beamformer is reconstructed as
$\widehat{\mathbf v}=\mathbf v(\widehat{\psi},\widehat{\phi})$.
For every realization of $(\widehat{\psi},\widehat{\phi})$, the reconstructed
beamforming vector satisfies
\[
\|\widehat{\mathbf v}\|_2^2
=
|\cos\widehat{\psi}|^2
+
|e^{j\widehat{\phi}}\sin\widehat{\psi}|^2
=
\cos^2\widehat{\psi}+\sin^2\widehat{\psi}
=
1,
\]
where the equality follows from the trigonometric identity
$\cos^2 x+\sin^2 x=1$ and the unit-modulus property
$|e^{j\widehat{\phi}}|=1$ for all $\widehat{\phi}\in[-\pi,\pi)$.
Thus, stochastic quantization of the angles preserves the unit-norm (and hence
orthonormality) constraint exactly, while affecting only the orientation of the
beamforming direction. The same reasoning extends to higher dimensions, where the beamforming matrix
is constructed as a product of Givens rotations and phase factors, each of
which remains unitary for all quantized angle values.

\subsection{Performance Guarantees}

We now characterize the bit error rate (BER) and achievable rate as a function of the proposed privacy-preserving perturbation applied to the beamforming matrix. The analysis proceeds by first mapping the resulting subspace distortion into an effective SNR, which directly governs both performance metrics. We start with the single stream case $N_{s} = 1$. The STA reconstructs a unit-norm
beamforming vector $\widehat{\mathbf v}\in\mathbb C^{N_t}$ from the compressed
CSI feedback (under the DP-SQ mechanism), and the received SNR is
\begin{align}
\gamma(\widehat{\mathbf v})
&\triangleq
\frac{P}{N_0}\, \cdot \|\mathbf H\widehat{\mathbf v}\|_2^2
\nonumber  \\
&=
\frac{P}{N_0}\, \cdot 
\widehat{\mathbf v}^{ H}\mathbf H^{ H}\mathbf H\,\widehat{\mathbf v}
\nonumber \\
&=
\frac{P}{N_0}\, \cdot 
\widehat{\mathbf v}^{ H}\mathbf V\boldsymbol{\Sigma}^2\mathbf V^{ H}\widehat{\mathbf v}
\label{eq:snr_svd}\\
&=
\frac{P}{N_0} \cdot
\sum_{i\ge 1}\sigma_i^2\, \cdot 
\big|\langle \mathbf v_i,\widehat{\mathbf v}\rangle\big|^2,
\label{eq:snr_expansion}
\end{align}
where \eqref{eq:snr_svd} follows from $\mathbf H^{ H}\mathbf H
=\mathbf V\boldsymbol{\Sigma}^2\mathbf V^{ H}$, and
\eqref{eq:snr_expansion} follows by expanding $\widehat{\mathbf v}$ in the
orthonormal basis $\{\mathbf v_i\}$. Now, let $\mathbf v^\star\triangleq \mathbf v_1$ denote the dominant right-singular
vector. Separating the dominant contribution yields
\begin{equation}
\gamma(\widehat{\mathbf v})
=
\frac{P}{N_0} \cdot \left(
\sigma_1^2\, \cdot \big|\langle \mathbf v^\star,\widehat{\mathbf v}\rangle\big|^2
+
\sum_{i\ge 2}\sigma_i^2\, \cdot \big|\langle \mathbf v_i,\widehat{\mathbf v}\rangle\big|^2
\right).
\label{eq:snr_split}
\end{equation}
Since $\widehat{\mathbf v}$ is unit norm and $\{\mathbf v_i\}$ is orthonormal, we have
\begin{equation}
\sum_{i\ge 1}\big|\langle \mathbf v_i,\widehat{\mathbf v}\rangle\big|^2 = 1,
\qquad
\sum_{i\ge 2}\big|\langle \mathbf v_i,\widehat{\mathbf v}\rangle\big|^2
=
1-\big|\langle \mathbf v^\star,\widehat{\mathbf v}\rangle\big|^2.
\label{eq:energy_conservation}
\end{equation}
For rank-one beamforming, the squared chordal distance satisfies
\begin{equation}
d_{\mathrm{chord}}^2(\mathbf v^\star,\widehat{\mathbf v})
=
1-\big|\langle \mathbf v^\star,\widehat{\mathbf v}\rangle\big|^2.
\label{eq:chordal_rank1}
\end{equation}

Using the fact that $\sigma_i^2\ge \sigma_2^2$ for all $i\ge 2$ and \eqref{eq:energy_conservation},
we further get
\begin{align}
\sum_{i\ge 2}\sigma_i^2\, \cdot \big|\langle \mathbf v_i,\widehat{\mathbf v}\rangle\big|^2
&\ge
\sigma_2^2 \cdot 
\sum_{i\ge 2}\big|\langle \mathbf v_i,\widehat{\mathbf v}\rangle\big|^2 =
\sigma_2^2\, \cdot  d_{\mathrm{chord}}^2(\mathbf v^\star,\widehat{\mathbf v}). \nonumber 
\end{align}
To this end, we obtain the following lower bound:
\begin{align}
\gamma(\widehat{\mathbf v})
\ \ge\
\gamma^\star(\mathbf H)\bigl(1-d_{\mathrm{chord}}^2(\mathbf v^\star,\widehat{\mathbf v})\bigr)
+
\gamma_2(\mathbf H)\,d_{\mathrm{chord}}^2(\mathbf v^\star,\widehat{\mathbf v}), \nonumber
\end{align}
where
\[
\gamma^\star(\mathbf H)\triangleq \frac{P}{N_0} \cdot \sigma_1^2(\mathbf H),
\qquad
\gamma_2(\mathbf H)\triangleq \frac{P}{N_0} \cdot \sigma_2^2(\mathbf H).
\]
In particular, since $\gamma_2(\mathbf H)\ge 0$, a simpler (but slightly looser)
bound is obtained by discarding the second term:
\begin{equation}
\gamma(\widehat{\mathbf v})
\ \ge\
\gamma^\star(\mathbf H)\bigl(1-d_{\mathrm{chord}}^2(\mathbf v^\star,\widehat{\mathbf v})\bigr)
\ \triangleq\
\gamma_{\mathrm{LB}}(\mathbf H).
\label{eq:snr_lb_simple_rank1}
\end{equation}

\subsubsection{Ergodic Achievable Rate}
\label{subsec:ergodic_rate_dp_sq}

We now characterize the achievable rate degradation induced solely by the
proposed DP-SQ mechanism. The instantaneous achievable rate (in bits-per-second-per Hz) is given by
\begin{align}
R
\triangleq
\log_2 \bigl(1+\gamma(\widehat{\mathbf v})\bigr),
\nonumber
\end{align}
Since the main source of randomness we are interested in, within a coherence block is the DP-SQ
mechanism, we define
\begin{equation}
\bar R
\triangleq
\mathbb E_{\mathrm{DP}}
\!\left[
\log_2 \bigl(1+\gamma(\widehat{\mathbf v})\bigr)
\right].
\label{eq:avg_rate_def}
\end{equation}
This corresponds to the achievable rate averaged over the stochastic feedback
mechanism for a fixed channel realization. If $\mathbf H$ varies ergodically
across blocks, an additional expectation over $\mathbf H$ yields the classical
ergodic rate \cite{el2011network}.

From Section~IV-A, the received SNR satisfies the affine lower bound
\begin{align}
\gamma(\widehat{\mathbf v})
\ \ge\
\gamma^\star(1-d)+\gamma_2 d,
\label{eq:snr_affine}
\end{align}
 where $d \triangleq d_{\mathrm{chord}}^2(\mathbf v^\star,\widehat{\mathbf v})
\in [0,1]$. While \eqref{eq:snr_affine} is tight, the resulting rate function is concave in
$d$, which prevents a direct Jensen-type lower bound on $\bar R$.

\vspace{0.3em}
\noindent\textbf{Smooth SNR lower bound.}
To enable tractable averaging, we introduce the conservative bound
\begin{equation}
\gamma(\widehat{\mathbf v})
\ \ge\
\frac{\gamma^\star}{1+\eta d},
\label{eq:snr_smooth_lb}
\end{equation}
where $\eta>0$ is chosen such that
\(
\eta \ge \frac{\gamma^\star-\gamma_2}{\gamma^\star}.
\)
This form is tight at $d=0$ and upper-bounds the worst-case SNR degradation
induced by subspace mis-alignment. We next present a lower bound guarantee on the achievable rate.

\vspace{0.5em}
\begin{theorem}[Ergodic rate lower bound under DP-SQ]
\label{thm:rate_dp_sq}
Let $\widehat{\mathbf v}$ be generated via DP-SQ and define
$d=d_{\mathrm{chord}}^2(\mathbf v^\star,\widehat{\mathbf v})$. Then the average
achievable rate satisfies
\begin{equation}
\bar R
\ \ge\
\log_2\!\left(
1+\frac{\gamma^\star}{1+\eta\,\mathbb E_{\mathrm{DP}}[d]}
\right).
\label{eq:avg_rate_lb_jensen}
\end{equation}
\end{theorem}

\begin{proof}
Substituting \eqref{eq:snr_smooth_lb} into \eqref{eq:avg_rate_def} yields
\begin{align}
\bar R
\ \ge\
\mathbb E_{\mathrm{DP}}
\!\left[
\log_2\!\left(1+\frac{\gamma^\star}{1+\eta d}\right)
\right].
\nonumber 
\end{align}
The function
\(
g(d)=\log_2\!\left(1+\frac{\gamma^\star}{1+\eta d}\right)
\)
is convex for $d\ge 0$. Applying Jensen's inequality completes the proof.
\end{proof}

From Theorem~1, the DP-SQ mechanism ensures
\(
\mathbb E_{\mathrm{DP}}[d]\le \mathsf D(\epsilon_\psi,\epsilon_\phi).
\)
Substituting into \eqref{eq:avg_rate_lb_jensen} yields
\begin{equation}
\bar R
\ \ge\
\log_2\!\left(
1+\frac{\gamma^\star}{1+\eta\,\mathsf D(\epsilon_\psi,\epsilon_\phi)}
\right).
\label{eq:rate_eps_final_exact}
\end{equation}

The above lower bound establishes an explicit
\emph{privacy--rate tradeoff}: tighter privacy budgets increase the expected
subspace distortion $\mathbb E[d]$, which reduces the effective SNR through the
factor $(1+\eta d)^{-1}$ and consequently degrades the achievable rate. Unlike
the BER metric, this degradation is logarithmic in SNR, implying that rate is
more robust to moderate privacy-induced perturbations, especially in the
high-SNR regime.

\subsubsection{Bit Error Rate}
\label{subsec:ber_analysis}

\vspace{0.3em}

For square $M$-QAM with Gray mapping over an AWGN channel, the instantaneous
bit error rate (BER) admits the standard approximation \cite{goldsmith2005wireless}:
\begin{align}
P_b(\widehat{\mathbf v})
\ \approx\
c_M\, \cdot
Q\!\Bigg(\sqrt{\frac{3\,\gamma(\widehat{\mathbf v})}{M-1}}\Bigg),
\quad
c_M \triangleq \frac{4}{\log_2 M}\Big(1-\frac{1}{\sqrt{M}}\Big),
\label{eq:ber_mqam}
\end{align}

We next show the impact of the DP-SQ mechanism on the BER.

\vspace{0.5em}
\begin{lemma}[BER bound under DP-SQ]
\label{lemma:ber_dp_sq}
Let $\widehat{\mathbf v}$ be the beamforming vector obtained via DP-SQ, and
let $d = d_{\mathrm{chord}}^2(\mathbf v^\star,\widehat{\mathbf v})$. Then, for
square $M$-QAM with Gray mapping, the instantaneous BER satisfies
\begin{equation}
P_b(\widehat{\mathbf v})
\ \le\
c_M\, \cdot 
Q\!\Bigg(
\sqrt{\frac{3\,\gamma^\star (1-d)}{M-1}}
\Bigg).
\label{eq:ber_bound_theorem}
\end{equation}
\end{lemma}

\begin{proof}
The result follows by substituting the SNR lower bound into the monotone decreasing function
\eqref{eq:ber_mqam}.
\end{proof}

\vspace{0.5em}
\noindent\textbf{Average BER under DP-SQ randomness.}
Since the DP-SQ mechanism induces randomness in $\widehat{\mathbf v}$, the
relevant performance metric is the average BER
\begin{equation}
\bar{P}_b
\ \triangleq\
\mathbb{E}_{\mathrm{DP}}\!\left[P_b(\widehat{\mathbf v})\right].
\end{equation}
Using Lemma~\ref{lemma:ber_dp_sq}, we obtain the bound
\begin{equation}
\bar{P}_b
\ \le\
c_M\, \cdot 
\mathbb{E}\!\left[
Q\!\Bigg(
\sqrt{\frac{3\,\gamma^\star (1-d)}{M-1}}
\Bigg)
\right],
\label{eq:avg_ber_general}
\end{equation}
where the expectation is taken over the distribution of the chordal distortion
$d$ induced by DP-SQ. Using the convexity of $Q(\sqrt{x})$ for $x>0$ and Jensen's inequality, we
further obtain
\begin{equation}
\bar{P}_b
\ \le\
c_M\, \cdot 
Q\!\Bigg(
\sqrt{\frac{3\,\gamma^\star (1-\mathbb{E}[d])}{M-1}}
\Bigg).
\label{eq:avg_ber_jensen}
\end{equation}

Equations~\eqref{eq:ber_bound_theorem}-\eqref{eq:avg_ber_jensen} establish a
direct and interpretable link between privacy and reliability. The DP-SQ
mechanism increases the expected subspace distortion $\mathbb{E}[d]$, which
reduces the effective SNR through a multiplicative factor $(1-d)$ and results
in an exponential degradation in BER via the $Q$-function. This characterization
is particularly sharp for high-order constellations, where reliability is
highly sensitive to SNR loss.

\subsection{Global DP via Full-Support Randomization}

The DP-SQ mechanism introduced above provides \emph{local} indistinguishability within each quantization interval, but does not satisfy \emph{global} DP (c.f. Definition~\ref{def:DP_beamforming}) over the entire angular domain due to its two-point output support. In particular, inputs from different quantization cells may induce output distributions with disjoint supports.

To address this limitation, we introduce a globally private extension that assigns non-zero probability to all reconstruction levels while preserving locality around the true angle. We refer to this globally private extension as \emph{DP-GSQ}, as it combines stochastic quantization with a full-support geometric randomization kernel.

\noindent \textbf{DP-GSQ Mechanism.}
Let $\mathcal Q=\{q_1,\dots,q_L\}$ denote the uniform quantization grid. For a given input angle $a$, let $q_i \le a \le q_{i+1}$ denote the bracketing levels. Define the stochastic interpolation weights
\[
\lambda_i(a)=\frac{q_{i+1}-a}{q_{i+1}-q_i}, 
\qquad
\lambda_{i+1}(a)=\frac{a-q_i}{q_{i+1}-q_i}.
\]

We define a truncated geometric kernel centered at level $q_j$ as follows:
\[
G_\tau(k \mid j)
=
\frac{(1-\tau)\,\tau^{d(k,j)}}{Z_j(\tau)}, 
\qquad k\in\{1,\dots,L\},
\]
where $0<\tau<1$, $d(k,j)$ is the index distance (or circular distance for $\phi$), and $Z_j(\tau)$ is a normalization constant. The globally private stochastic quantizer releases $\widehat a \in \mathcal Q$ according to
\begin{equation}
\Pr(\widehat a = q_k \mid a)
=
\lambda_i(a)\, G_\tau(k \mid i)
+
\lambda_{i+1}(a)\, G_\tau(k \mid i+1).
\label{eq:global_dpsq}
\end{equation}

\begin{figure}[t]
    \centering
    \includegraphics[width=0.95\columnwidth]{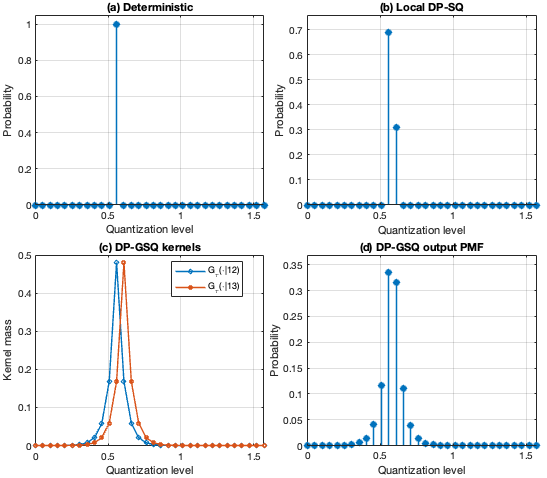}
    \caption{\small
    Visualization of deterministic quantization, local DP-SQ, and the proposed globally private DP-GSQ mechanism for a fixed input angle.
    (a) Deterministic quantization places all probability mass on a single reconstruction level.
    (b) Local DP-SQ randomizes only between the two neighboring levels.
    (c) DP-GSQ uses truncated geometric kernels centered at the two bracketing levels.
    (d) The final DP-GSQ output distribution is a convex combination of these kernels and assigns non-zero probability to all quantization levels.
    This full-support property eliminates zero-probability outputs and enables a global differential privacy guarantee over the entire angular domain.
    }
    \label{fig:dpgsq_kernel}
\end{figure}

Fig.~\ref{fig:dpgsq_kernel} illustrates the key distinction between the two mechanisms. While local DP-SQ perturbs only within the bracketing quantization cell, DP-GSQ ($\tau = 0.35$) spreads a geometrically decaying tail over the entire angle codebook. This full-support randomization is precisely what enables a finite global DP guarantee.

Since $G_\tau(k\mid j) > 0$ for all $k,j$, every output level has non-zero probability under any input. Therefore, the mechanism satisfies $\varepsilon$-local DP over the full angular domain.

\begin{theorem}[Global DP guarantee]
The mechanism in \eqref{eq:global_dpsq} satisfies $\varepsilon_a$-local DP, i.e.,
\[
\Pr(\widehat a \in S \mid a)
\le
e^{\varepsilon_a}\Pr(\widehat a \in S \mid a'),
\quad \forall a,a',\ \forall S\subseteq \mathcal Q,
\]
where
\[
\varepsilon_a
\le
\log \left(
\frac{\max_{j,k} G_\tau(k\mid j)}
{\min_{j,k} G_\tau(k\mid j)}
\right).
\]
\end{theorem}

\noindent \textbf{Proof Sketch.}
The proof follows by bounding the likelihood ratio
\[
\sup_{a,a'} \sup_{k}
\frac{\Pr(\widehat a = q_k \mid a)}
{\Pr(\widehat a = q_k \mid a')}.
\]
Since $\Pr(\widehat a = q_k \mid a)$ is a convex combination of strictly positive kernels $G_\tau(k\mid i)$ and $G_\tau(k\mid i+1)$, the ratio is bounded by the ratio between the maximum and minimum values of $G_\tau(k\mid j)$ across all indices. This yields the stated bound.

\subsubsection{Operational interpretation of DP-GSQ}

The kernel parameter $\tau$ controls the concentration of the output distribution over the quantization alphabet. In particular, the truncated geometric kernel assigns probability mass according to $G_\tau(k\mid j)\propto \tau^{d(k,j)}$, so that $\tau$ acts as a \emph{density-flattening parameter}. In the regime $\tau \to 0$, the kernel becomes sharply concentrated around the bracketing reconstruction levels, and the DP-GSQ mechanism behaves similarly to a localized stochastic quantizer with negligible tails. In contrast, as $\tau \to 1$, the geometric decay vanishes and the output distribution approaches a uniform distribution over all quantization levels. Consequently, DP-GSQ continuously interpolates between localized stochastic quantization and full-support randomization, providing a tunable mechanism for controlling the privacy--utility tradeoff.

As $\tau$ increases, the output distribution becomes more uniform, reducing the distinguishability between different inputs and thereby decreasing $\varepsilon_{a}(\tau)$, $a \in \{\theta, \phi\}$. Conversely, smaller values of $\tau$ concentrate probability mass near the true angle, increasing distinguishability and hence the privacy leakage. This establishes a direct and interpretable relationship between the kernel parameter $\tau$, the induced privacy level, and the resulting distortion.

\noindent \textbf{Distortion analysis.}
Under the proposed mechanism, the per-angle distortion is given by
\[
\mathbb E\!\left[(\widehat a - a)^2 \mid a\right]
=
\sum_{k=1}^L (q_k - a)^2\,\Pr(\widehat a = q_k \mid a).
\]
Averaging over $a$ within each quantization interval yields the effective angular distortion
\[
\sigma_{a,G}^2
=
\frac{1}{\Delta}
\int_{q_i}^{q_{i+1}}
\sum_{k=1}^L (q_k - a)^2\,\Pr(\widehat a = q_k \mid a)\, da.
\]

\begin{remark}
Compared to the original DP-SQ mechanism, the globally private extension ensures rigorous DP over the entire angular domain at the cost of increased distortion, as probability mass is spread across all quantization levels. This highlights a fundamental tradeoff between global privacy guarantees and beamforming accuracy.
\end{remark}

\begin{proposition}[MSE bound for DP--GSQ]
Consider the globally private stochastic quantization mechanism in \eqref{eq:global_dpsq} over a uniform grid $\mathcal Q=\{q_1,\dots,q_L\}$ with spacing $\Delta$. Define
\[
M_\tau(j)\triangleq \sum_{k=1}^L (q_k-q_j)^2 G_\tau(k\mid j).
\]
Then, for any input $a\in[q_i,q_{i+1}]$, the conditional distortion satisfies
\[
\mathbb E\!\left[(\widehat a-a)^2\mid a\right]
\le
2\max_{1\le j\le L} M_\tau(j)+\frac{\Delta^2}{2}.
\]
Moreover, if $a$ is uniformly distributed within its quantization interval, then
\[
\mathbb E\!\left[(\widehat a-a)^2\right]
\le
2\max_{1\le j\le L} M_\tau(j)+\frac{\Delta^2}{6}.
\]
\end{proposition}

\begin{proof}
By definition,
\[
\Pr(\widehat a=q_k\mid a)
=
\lambda_i(a)G_\tau(k\mid i)+\lambda_{i+1}(a)G_\tau(k\mid i+1).
\]
Hence
\begin{align}
\mathbb E[(\widehat a-a)^2\mid a]
=
\lambda_i(a)\sum_k (q_k-a)^2G_\tau(k\mid i) \nonumber \\
+
\lambda_{i+1}(a)\sum_k (q_k-a)^2G_\tau(k\mid i+1). \nonumber 
\end{align}
Using
\[
(q_k-a)^2 \le 2(q_k-q_j)^2+2(q_j-a)^2,
\]
for $j\in\{i,i+1\}$, we obtain
\[
\sum_k (q_k-a)^2G_\tau(k\mid j)
\le
2M_\tau(j)+2(q_j-a)^2.
\]
Substituting this into the previous expression gives
\begin{align}
&\mathbb E[(\widehat a-a)^2\mid a]
\le
2\max_j M_\tau(j) \nonumber \\
&+
2\bigl(\lambda_i(a)(a-q_i)^2+\lambda_{i+1}(a)(q_{i+1}-a)^2\bigr).
\end{align}
Since the last term is upper bounded by $\Delta^2/2$, the conditional bound follows. Averaging over $a$ uniformly within the interval yields the second statement.
\end{proof}

\begin{corollary}[Subspace distortion under DP-GSQ]
Let $\sigma_{\psi,\mathrm{G}}^2 \triangleq \mathbb E[(\widehat\psi-\psi)^2]$ and
$\sigma_{\phi,\mathrm{G}}^2 \triangleq \mathbb E[(\widehat\phi-\phi)^2]$
denote the per-angle mean squared errors induced by the DP-GSQ mechanism.
Then the expected squared chordal distance between the ideal subspace
$\mathcal S^\star$ and the reconstructed subspace
$\widehat{\mathcal S}_{\mathrm{G}}$ satisfies
\[
\mathbb E\!\left[d_{\mathrm{chord}}^2(\mathcal S^\star,\widehat{\mathcal S}_{\mathrm{G}})\right]
\le
d_q^2 + 2 \cdot N_s \cdot N_{\mathrm{tot}} \cdot
\bigl(\sigma_{\psi,\mathrm{G}}^2+\sigma_{\phi,\mathrm{G}}^2\bigr).
\]
\end{corollary}

The result follows by applying the same Givens-cascade perturbation argument as in Theorem~\ref{thm:dpsq-utility-analytic-general-Ns}, with the per-angle error variances replaced by those induced by the DP-GSQ mechanism. The corresponding achievable rate and BER characterizations follow along the same lines.

\section{Numerical Simulations}
\label{sec:numerical_results}

In this section, we evaluate the impact of DP Givens angle
quantization scheme on both downlink beamforming utility and the adversary’s
ability to estimate user motion speed from the reported CSI angular parameters.  The downlink channel $\mathbf{H}(t)$ follows  a standard time-varying multipath fading model at each coherence block, the receiver computes the right-singular
vector of $\mathbf{H}(t)$ and encodes it into Givens rotation and
phase angles following the 802.11 compressed beamforming format.
We apply our DP stochastic quantizer to the true angles, drawing
each reported angle $\widehat{\psi}$ or $\widehat{\phi}$ from an
$\varepsilon$-DP distribution centered at its true value.  The AP uses
the privatized angles to reconstruct a semi-unitary transmit beamformer $\widehat{\mathbf{V}}$, while a
passive adversary receives the same angles and attempts speed
estimation following the micro-Doppler extraction pipeline in \cite{cominelli2024physical}.  \\

\begin{figure}[t]
    \centering

    \begin{subfigure}{0.48\columnwidth}
        \centering
        \includegraphics[width=\linewidth]{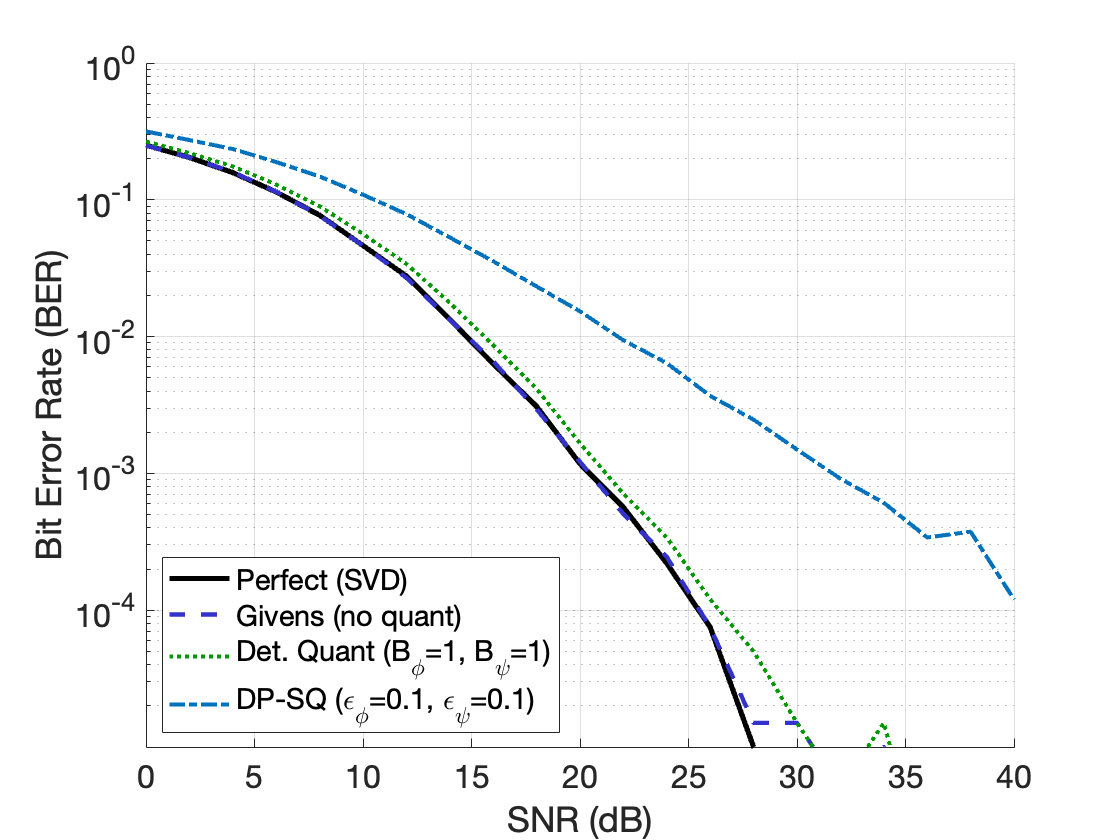}
        \caption{\small{16-QAM, 1-bit feedback}}
        \label{fig:ber_16qam_1bit}
    \end{subfigure}
    \hfill
    \begin{subfigure}{0.48\columnwidth}
        \centering
        \includegraphics[width=\linewidth]{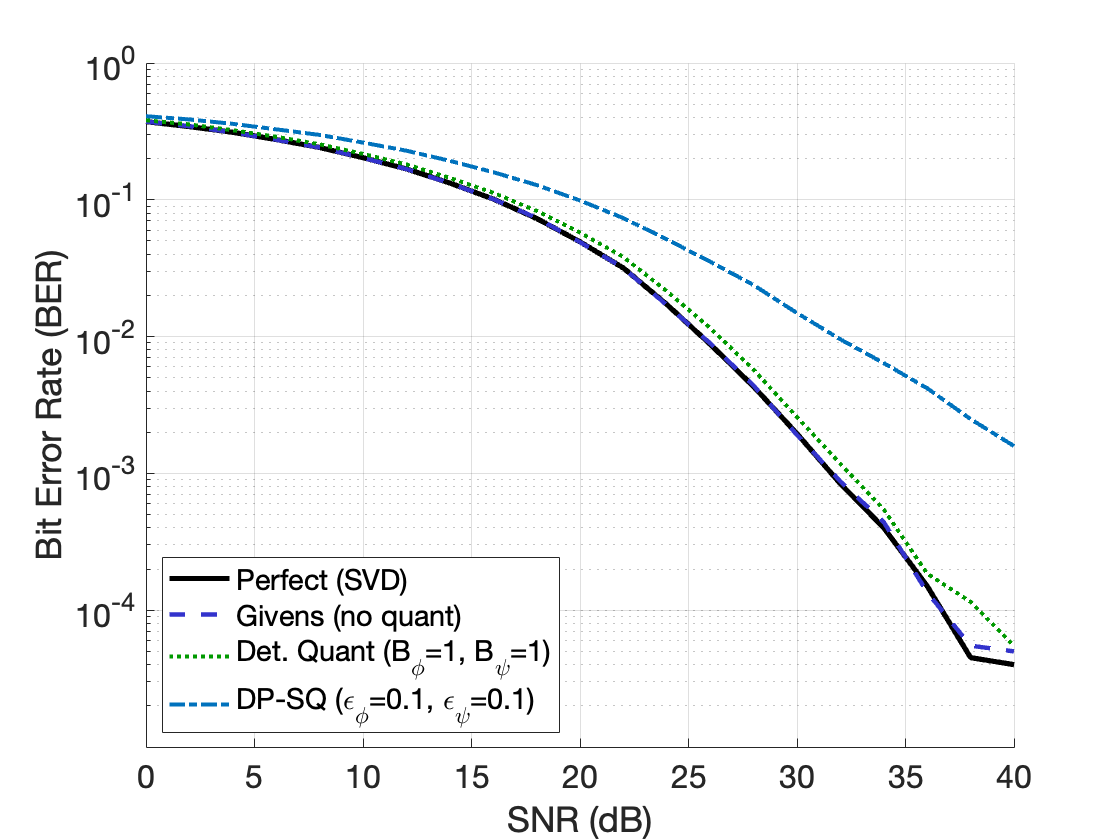}
        \caption{\small{256-QAM, 1-bit feedback}}
        \label{fig:ber_256qam_1bit}
    \end{subfigure}

    \vspace{6pt}

    \begin{subfigure}{0.48\columnwidth}
        \centering
        \includegraphics[width=\linewidth]{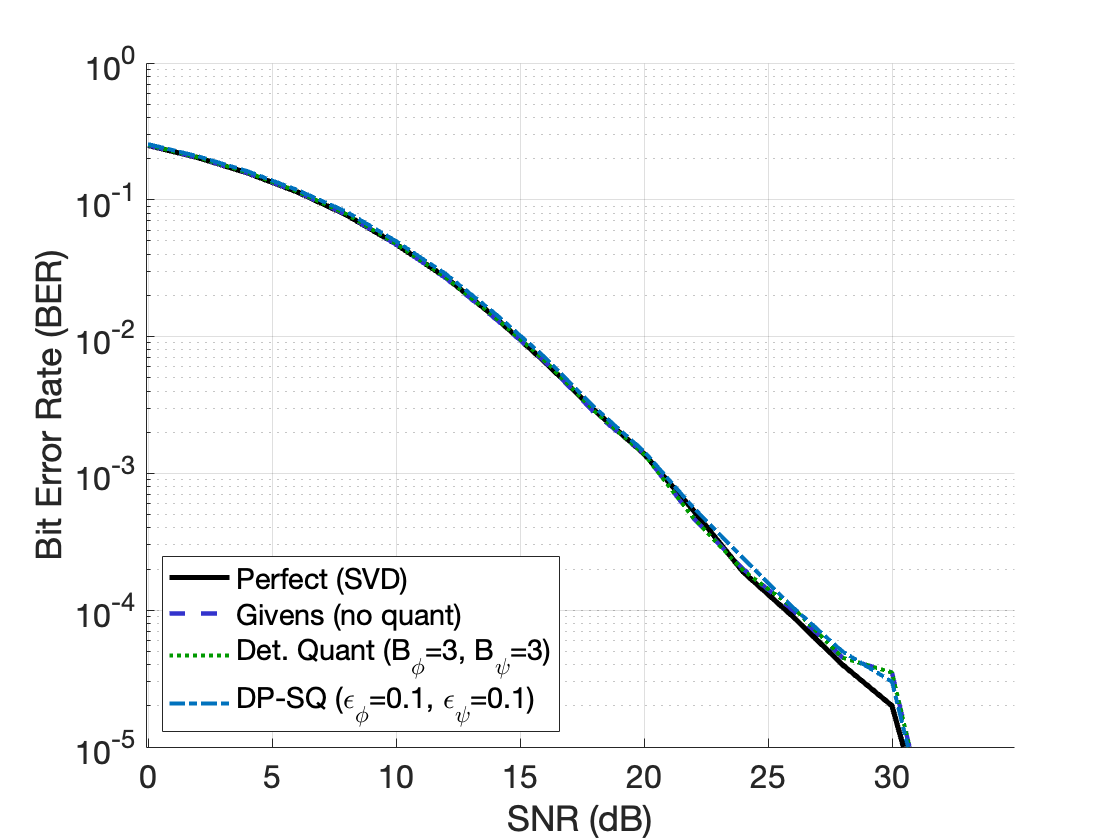}
        \caption{\small{16-QAM, 3-bit feedback}}
        \label{fig:ber_16qam_3bit}
    \end{subfigure}
    \hfill
    \begin{subfigure}{0.48\columnwidth}
        \centering
        \includegraphics[width=\linewidth]{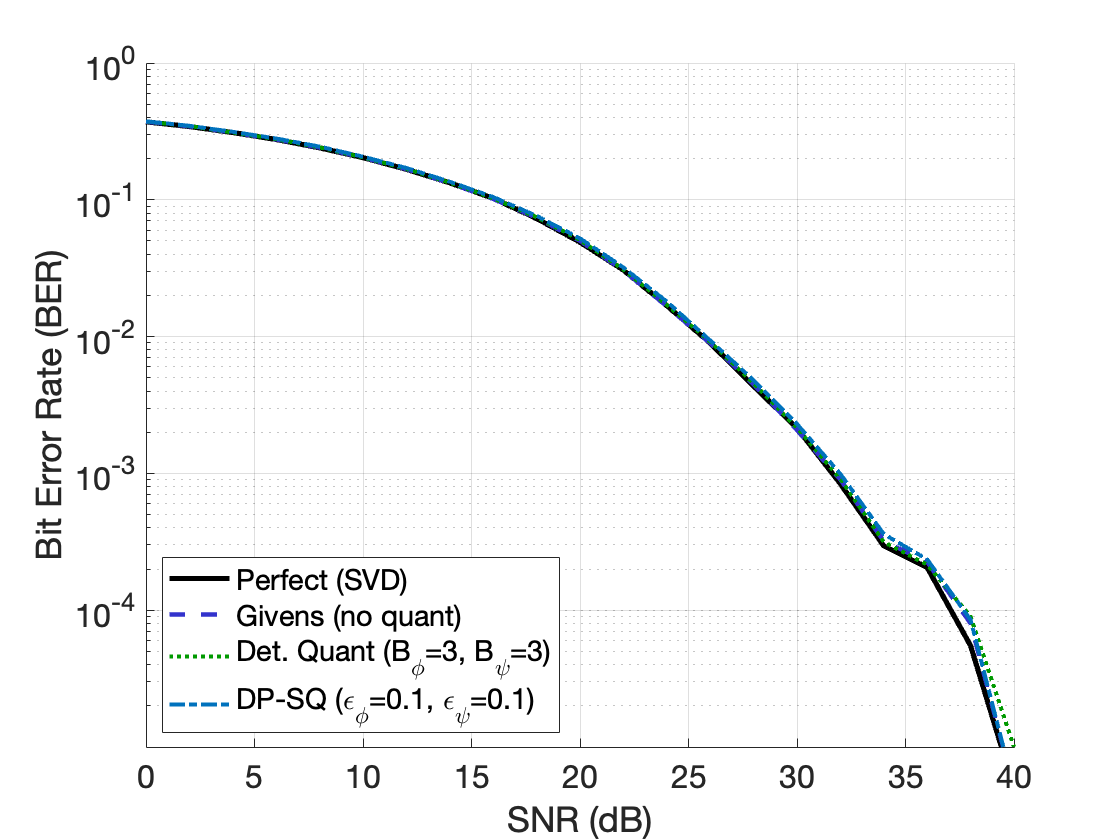}
        \caption{\small{256-QAM, 3-bit feedback}}
        \label{fig:ber_256qam_3bit}
    \end{subfigure}

    \caption{
        \small{
        Impact of privacy-preserving angle feedback on BER
        for different modulation orders and angle quantization resolutions where $\varepsilon = 0.1$.
        }
    }
    \label{fig:ber_vs_snr_all}
\end{figure}

\noindent \textbf{Impact on Bit Error Rate (BER).} In Fig.~\ref{fig:ber_vs_snr_all}, we report the impact of DP–SQ on link performance as measured by BER.
The first row corresponds to low-resolution (1-bit) CSI feedback, while the second row shows high-resolution (3-bit) feedback. As expected, higher modulation orders or lower angular resolutions increase the system’s sensitivity to DP-induced beam mis-alignment, resulting in elevated BER across SNR levels. Notably, in the high-resolution setting, the proposed DP–SQ mechanism achieves BER performance nearly indistinguishable from the non-private baseline, demonstrating that strong privacy can be attained with minimal loss when sufficient angular resolution is available. 

\begin{figure}[t]
    \centering
    \includegraphics[width= \linewidth]{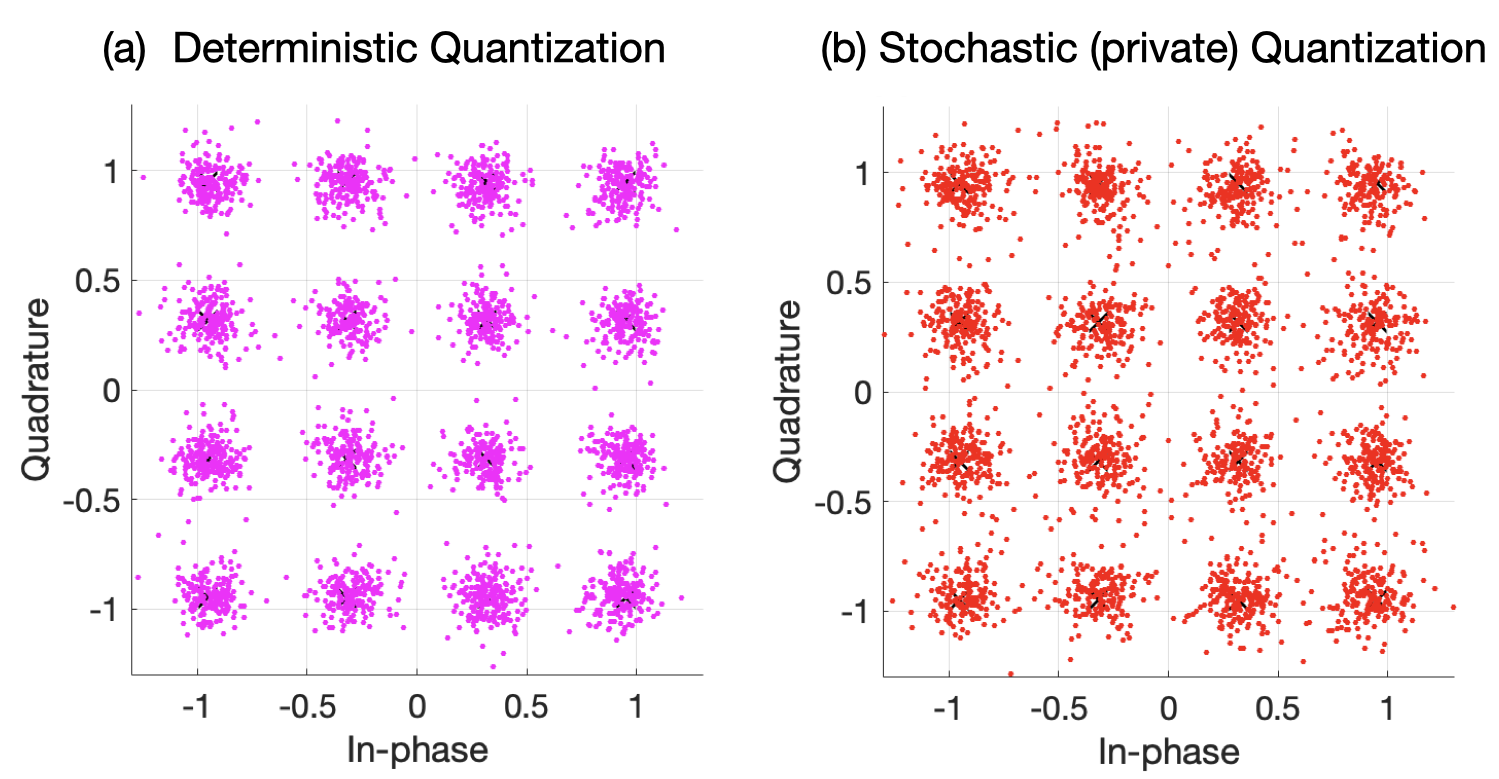}
    \caption{\small{Constellation comparison for a single spatial stream under 
(a) deterministic midpoint quantization of the Givens angles, and 
(b) the proposed differentially private stochastic quantization (DP-SQ) mechanism for $N_{t} = 2$, $N_{r} = 1$, $16$-QAM, $\operatorname{SNR} = 15 $ dB and $\varepsilon = 0.1$, $B_{\phi} = B_{\psi} = 1$ bit. 
}}
    \label{fig:constellations}
\end{figure}

In Fig.~\ref{fig:constellations}, we visualize the received constellation points to provide intuition behind these trends.
SVD-based beamforming produces tightly clustered decision points around the ideal 16-QAM constellation.
Deterministic quantization introduces moderate angular distortion, broadening the symbol clouds.
The DP–SQ mechanism injects controlled stochasticity into the angular representation, further dispersing the received symbols while offering formal DP guarantees. \\

\noindent \textbf{Impact on Beamforming-Gain.} To quantify the distortion introduced by the different privatization 
mechanisms, we evaluate a \emph{relative beamforming-gain} metric that 
captures how closely a privatized precoder preserves the energy projection
of the optimal transmitter.  For stream $k$ with beamforming vector $\mathbf{v}_k$ (the $k$-th column 
of $\mathbf{V}$), the resulting effective channel is
$\mathbf{h}_k^{\mathrm{eff}}
\,=\,\mathbf{H}\mathbf{v}_k$. The per-stream beamforming gain is defined as
$G_k
\, \triangleq \,{\|\mathbf{H}\mathbf{v}_k\|^{2}}
           /{\|\mathbf{H}\mathbf{v}_k^{*}\|^{2}}
\in[0,1]$,
and we report the average gain as $G \,=\, \frac{1}{N_s}\sum_{k=1}^{N_s} G_k$,
which equals $1$ for perfect (unperturbed) beamforming and decreases as
the angular perturbations introduced by quantization or the privacy
mechanism increase.

\begin{figure}[t]
    \centering
        \includegraphics[width= 0.85\linewidth]{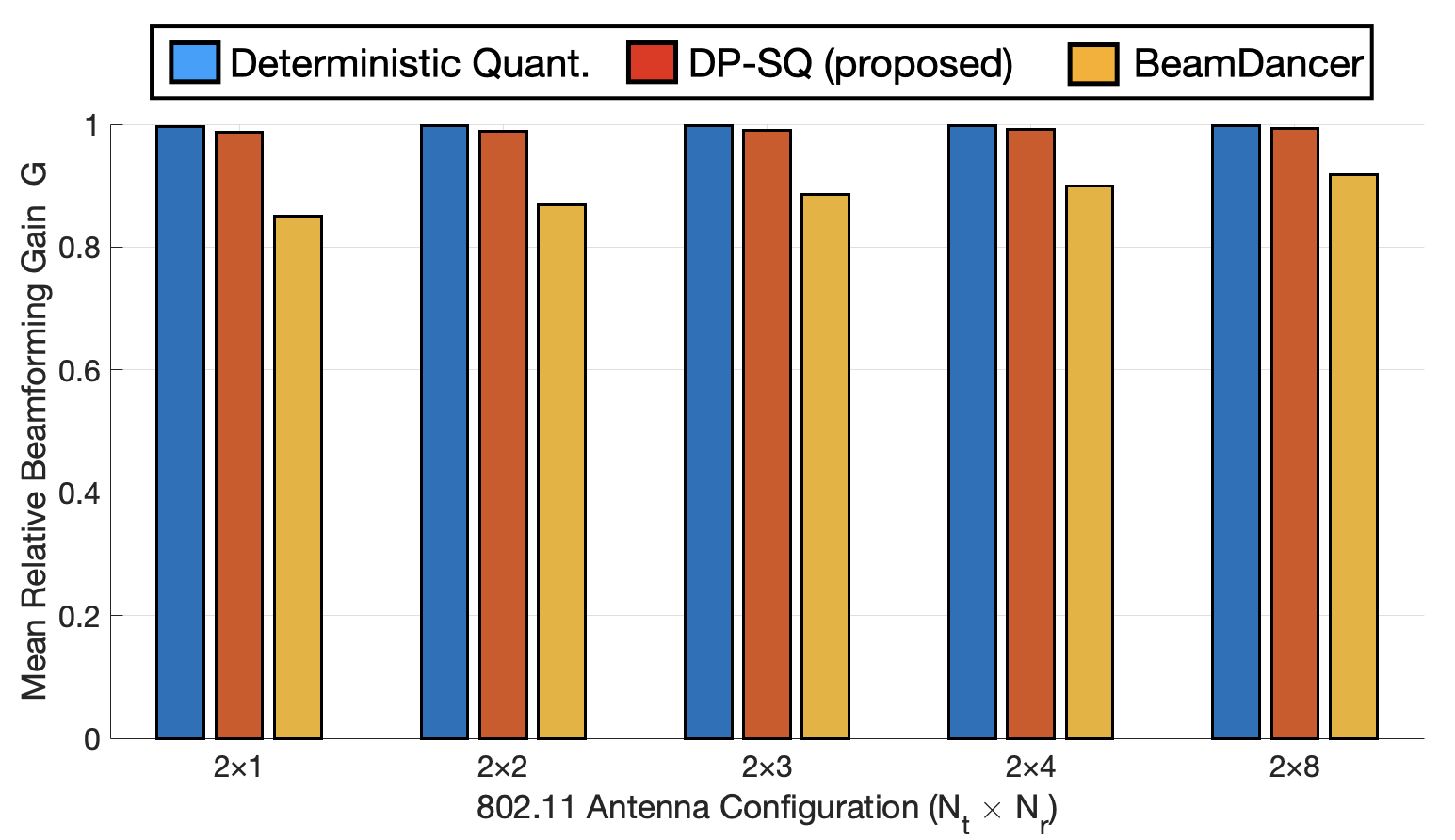}

    \caption{
    \small{
    Relative beamforming gain for several IEEE~802.11 antenna configurations 
    ($2\times1$, $2\times2$, $2\times3$, $2\times4$, $2\times8$).
    The perfect SVD beamformer defines the reference gain $G=1$, where $\epsilon_{\phi} = \epsilon_{\psi}  = 0.1$, $B_{\phi} = 6$ bits and $B_{\psi} = 3$ bits. For the BeamDancer scheme~\cite{cominelli2024physical}, we adopt the randomization parameters that ensure the adversary’s activity classification accuracy remains strictly below $50 \%$.
    }
    }
    \label{fig:beamforming_gain_subfig}
\end{figure}

Next, in Fig.~\ref{fig:beamforming_gain_subfig}, we report the average beamforming gain $G$ over $5000$ Rayleigh fading realizations for several IEEE~802.11 antenna configurations with $(N_t,N_r)\in{(2,1),(2,2),(2,3),(2,4),(2,8)}$ and a single spatial stream ($N_s=1$). Further,  Fig.~\ref{fig:beamforming_gain_subfig} reports the mean gain achieved by three schemes:
(i) deterministic mid-point quantization of the Givens angles,
(ii) the proposed differentially private stochastic quantization (DP–SQ), and
(iii) BeamDancer-style randomized beamforming \cite{cominelli2024physical}.
The perfect SVD beamformer provides the reference gain $G=1$. Deterministic quantization incurs moderate loss due to finite angular resolution. The proposed DP–SQ mechanism introduces controlled stochastic perturbations governed by the privacy parameters $(\varepsilon_\phi,\varepsilon_\psi)$, producing a tunable privacy–utility tradeoff: smaller values of $\varepsilon$ increase privacy but yield additional beam mis-alignment and corresponding gain reduction.
BeamDancer \cite{cominelli2024physical} exhibits the largest degradation, as its high-variance integer-bin perturbations cause substantial angular distortion. Across all antenna configurations, the trend is consistent: privacy-driven angular randomization directly impacts achievable array gain, with the effect magnified in larger arrays, yet the proposed DP-SQ scheme preserves compatibility with the 802.11 feedback format while offering significantly improved privacy–utility tradeoff. \\

\begin{table}[t]
\centering
\caption{\small{Default WiFi channel and system simulation parameters.}}
\label{tab:wifi_params}
\begin{tabular}{l l c}
\hline
\textbf{Parameter} & \textbf{Symbol} & \textbf{Value} \\
\hline
\multicolumn{3}{l}{\underline{\textit{MIMO Configuration}}} \\
Transmit antennas         & $N_t$         & 2              \\
Receive antennas          & $N_r$         & 1              \\
Spatial streams           & $N_s$         & 1              \\
Antenna spacing           & $d$           & $0.5\lambda$   \\
\hline
\multicolumn{3}{l}{\underline{\textit{OFDM / Bandwidth}}} \\
Carrier frequency         & $f_c$         & 5.785\,GHz     \\
Channel bandwidth         & $B$           & 20\,MHz        \\
Number of subcarriers     & $N_{sc}$      & 256            \\
Subcarrier spacing        & $\Delta f$    & 78.125\,kHz    \\
\hline
\multicolumn{3}{l}{\underline{\textit{Channel Model}}} \\
Rician $K$-factor         & $K$           & 5\,dB          \\
Number of multipath taps  & $L$           & 20             \\
Maximum excess delay      & $\tau_{\max}$ & 4 samples      \\
User direction of arrival & $\theta_0$    & $15^{\circ}$   \\
\hline
\multicolumn{3}{l}{\underline{\textit{Temporal Sampling}}} \\
Snapshot interval         & $\Delta t$    & 1\,ms          \\
Number of snapshots       & $T$           & 5000           \\
Observation duration      & $T\Delta t$   & 5\,s           \\
\hline
\end{tabular}
\end{table}

\begin{figure}[t]
\centering

\includegraphics[width= 0.85 \columnwidth]{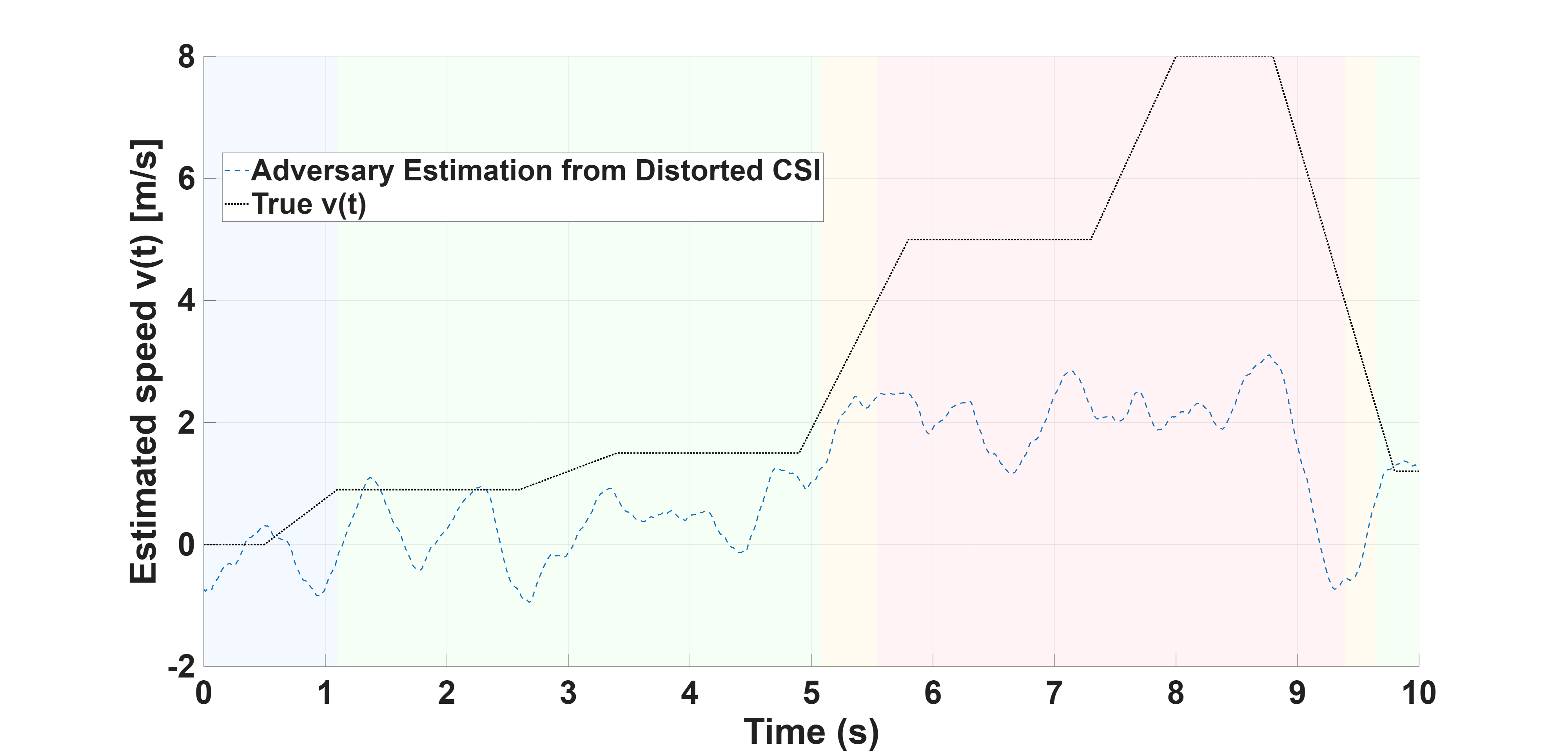}
\centering
    \caption{\small{Estimated speed with privacy-based perturbations and the adversary as the observer,  where $\epsilon_{\phi} = \epsilon_{\psi}  = 0.1$, $B_{\phi} = 6$ bits and $B_{\psi} = 3$ bits.}}
    \label{fig:privacy_example}
\end{figure}

\begin{figure}[t]
\centering
\begin{subfigure}[t]{0.5\textwidth}
\centering
\includegraphics[width= \textwidth]{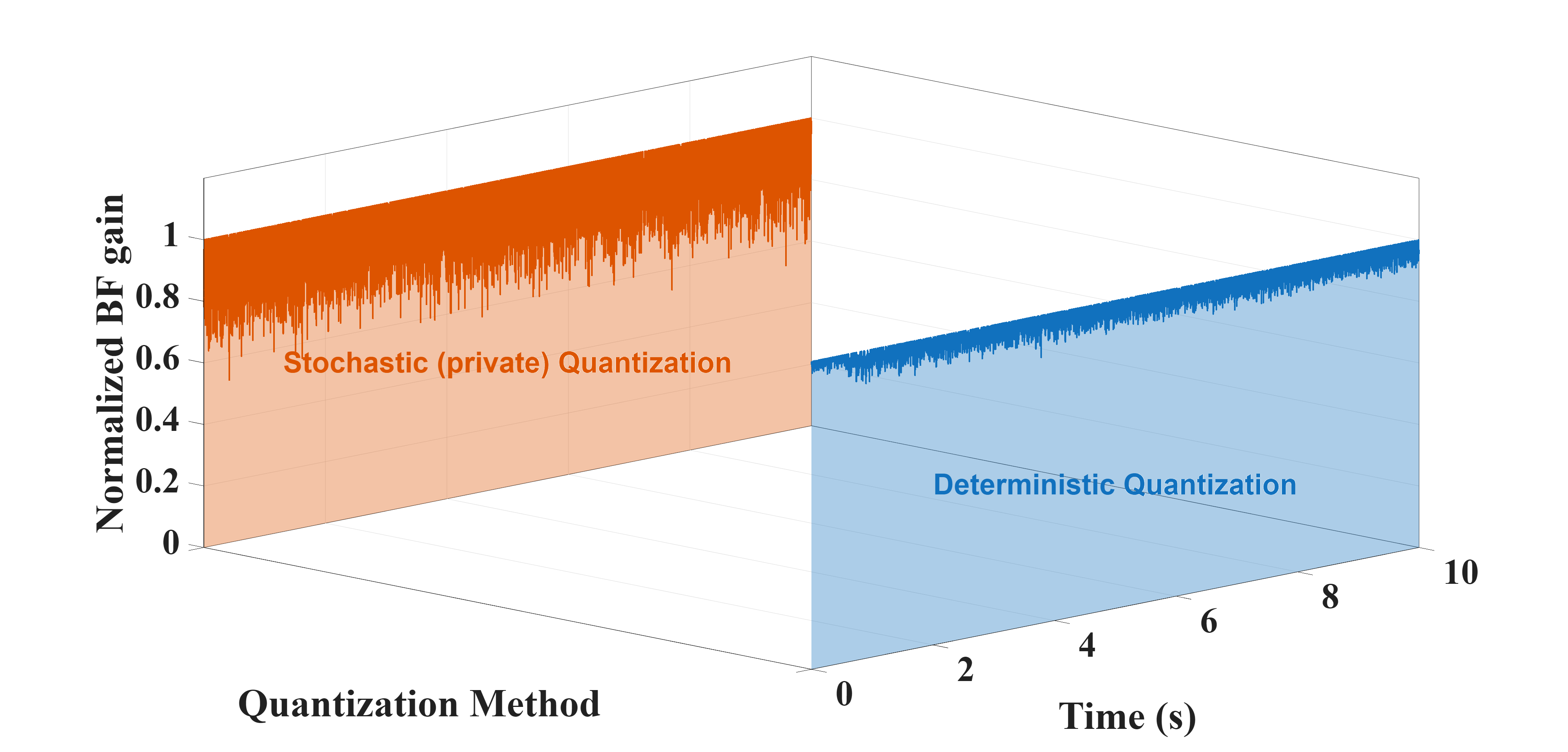}
\caption{}
\label{fig:BF_Gain}
\end{subfigure}
\begin{subfigure}[t]{0.5\textwidth}
\centering
\includegraphics[width= 0.85 \textwidth]{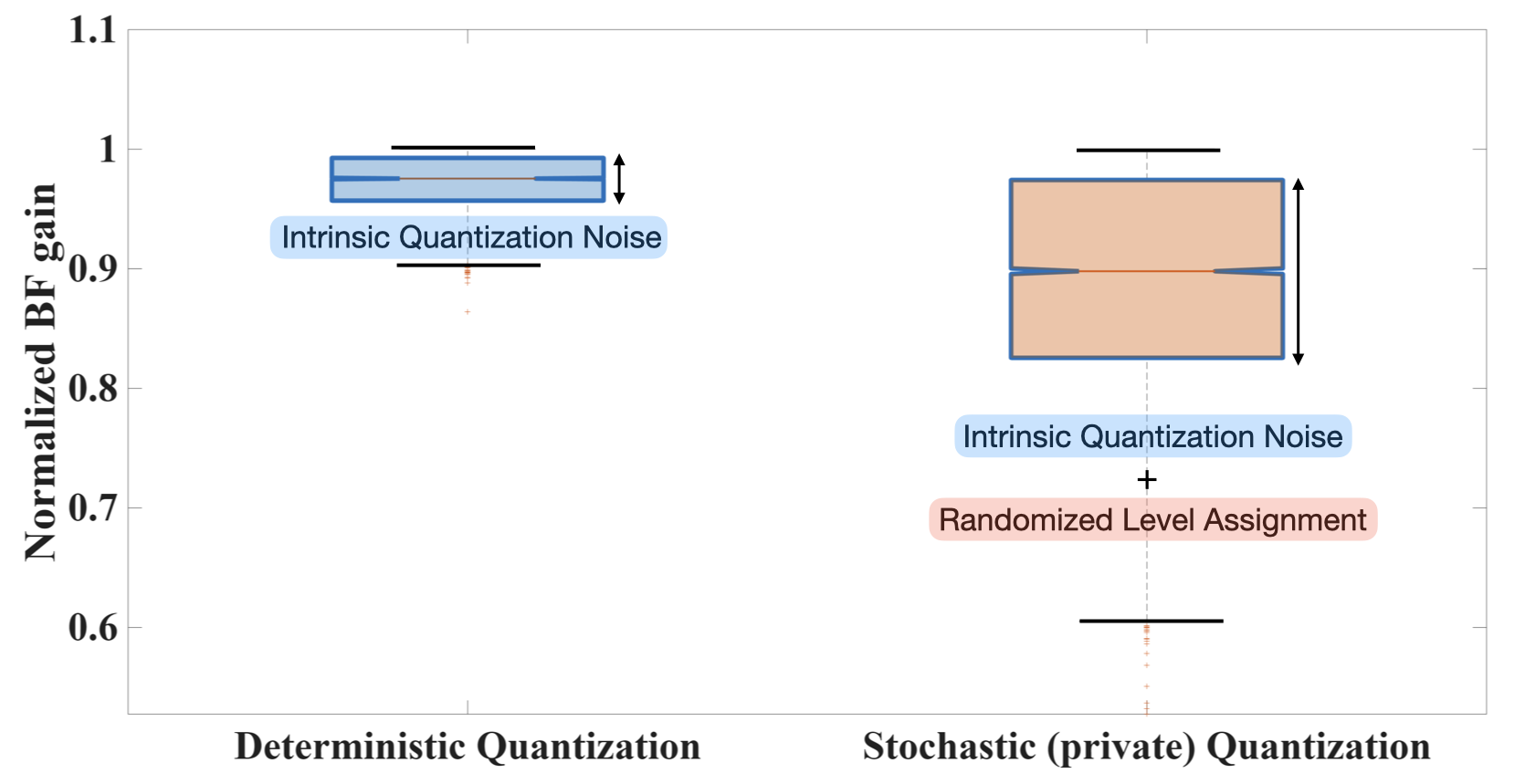}
\centering
\caption{}
\label{fig:Speed_Estimation_Distort}
\end{subfigure}
    \caption{\small{ Direct beamforming gain comparison   under deterministic (non-private) and stochastic (private) quantization. (a) Beamforming gain over time. (b) Distribution of the normalized per-stream beamforming gain. The proposed privacy mechanism disrupts an adversary’s ability to infer STA motion while preserving a near-optimal median beamforming gain and maintaining communication utility.}}
   \label{fig:BF_Gain_box}
\end{figure}

\begin{figure*}[t]
    \centering
    \begin{subfigure}[t]{0.326\textwidth}
        \centering
        \includegraphics[width=\linewidth]{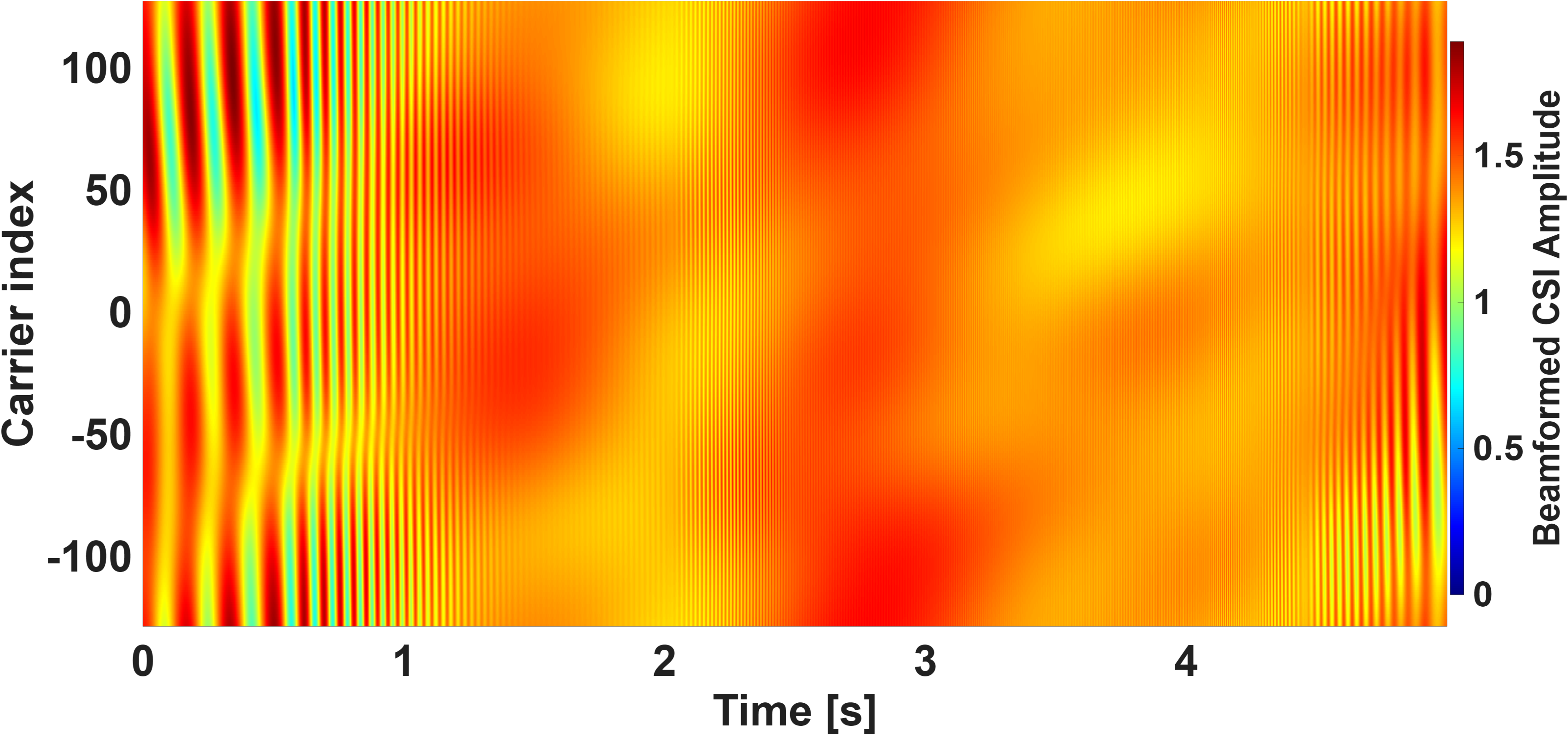}
        \caption{\small{CSI from Deterministic Quantization}}
        \label{fig:CSI_determine}
    \end{subfigure}\hspace{1mm}%
    \begin{subfigure}[t]{0.326\textwidth}
        \centering
        \includegraphics[width=\linewidth]{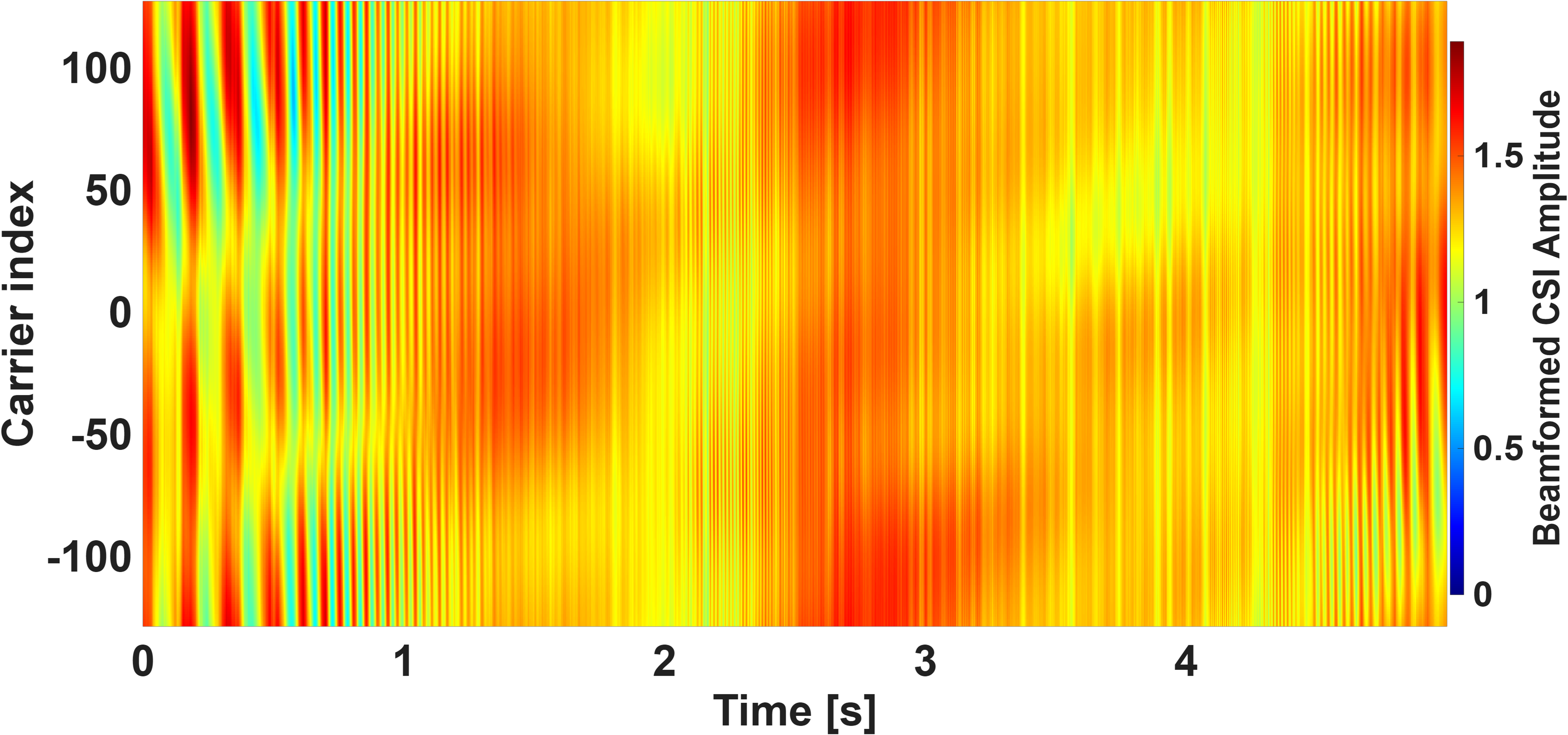}
        \caption{\small{CSI from DP-SQ}}
        \label{fig:CSI_DP}
    \end{subfigure}\hspace{1mm}%
    \begin{subfigure}[t]{0.326\textwidth}
        \centering
        \includegraphics[width=\linewidth]{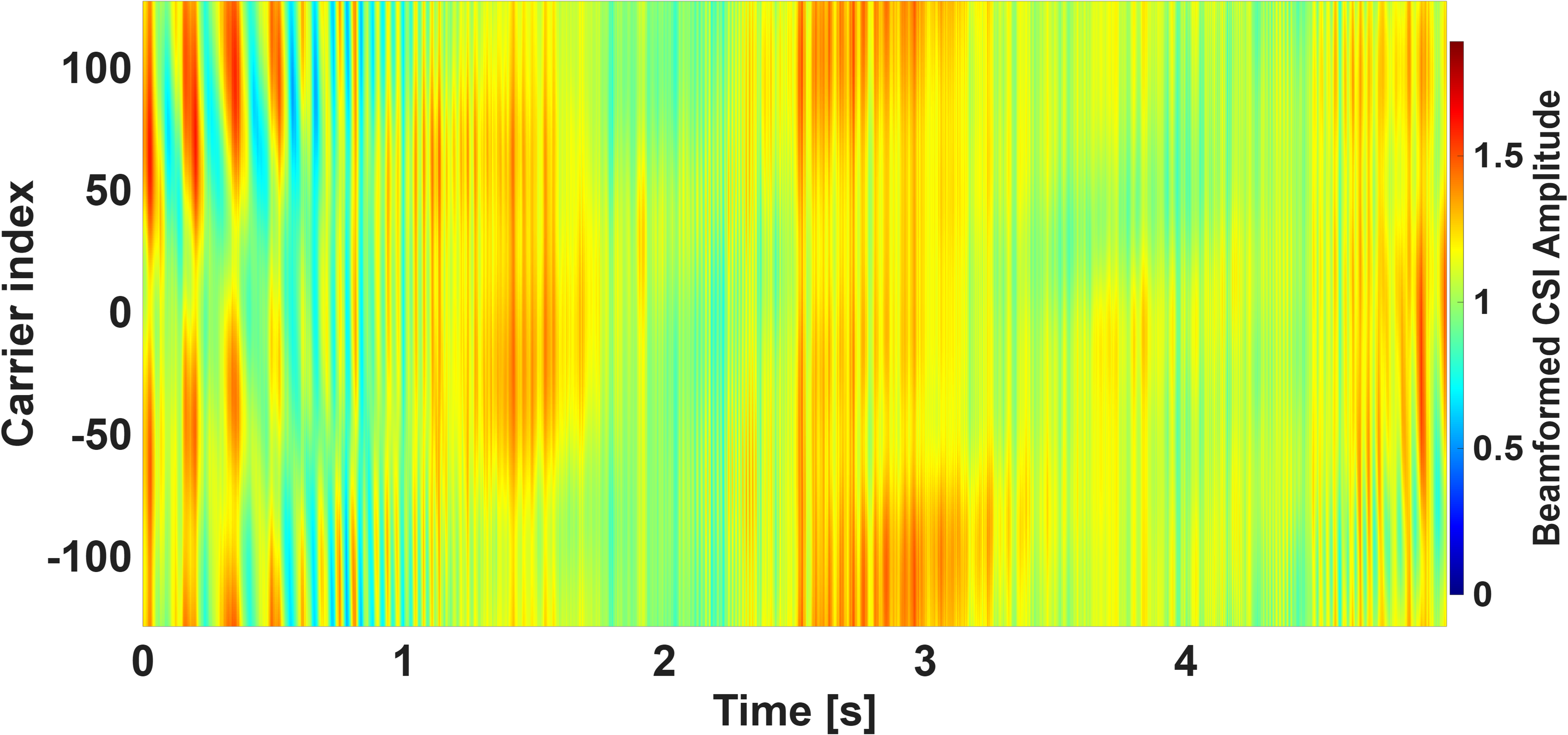}
        \caption{\small{CSI from DP--GSQ}}
        \label{fig:CSI_GDP}
    \end{subfigure}

    \caption{
        \small{
        Beamformed CSI amplitude across subcarrier index and time for a user transitioning through stationary, walking, jogging, and running activity zones.
        }
    }
    \label{fig:CSI_Heatmap}
\end{figure*}

\noindent \textbf{Robustness to Adversarial Estimation.} We further verify our ability to perturb the adversary's access to the legitimate user information. Table~\ref{tab:wifi_params} summarizes the simulation parameters. We model an  IEEE~802.11ac MIMO-OFDM link \cite{park2013ieee80211ac} at $f_c = 5.785$\,GHz with a $20$\,MHz bandwidth 
and $N_{sc} = 256$ subcarriers. The channel follows a Rician fading model ($K = 5$\,dB) with $L = 20$ multipath taps, where the dominant line-of-sight component carries the user's Doppler signature. CSI snapshots are sampled every  $1$\,ms over $5$ seconds, providing sufficient temporal resolution to capture the micro-Doppler shifts induced by human motion as the user transitions through 
stationary, walking, jogging, and running activity zones.
Fig.~\ref{fig:privacy_example} provides an example demonstration of the effectiveness of our proposed privacy-based perturbations over time: despite the injected noise on the compressed CSI, the normalized per-stream beamforming gain generally remains close. Fig.~\ref{fig:BF_Gain_box} further quantifies this, where the median gain obtained from the distorted CSI is $0.89$ (compared to the deterministic median gain of $0.97$ and never drops below  $0.52$. 
A closer look at the CSI, as shown in Fig.~\ref{fig:CSI_Heatmap} reveal how this occurs. Under deterministic quantization, the heatmap exhibits fine-grained frequency-selective fading patterns and rapid temporal fluctuations that encode micro-Doppler signatures correlated with user speed. However, under DP-SQ, these patterns are altered but remain well-correlated with the deterministic quantization. Thus, the overall beamforming gain is expected to be similar to the deterministic-quantization equivalent, and the speed estimation is expected to be degraded due to changes in the ground-truth temporal CSI structure. However, under DP-GSQ, the patterns are visibly distorted, obscuring the speed-dependent micro-Doppler structure while preserving the broadband channel envelope needed for beamforming.

\begin{figure*}[t]
    \centering
    \begin{subfigure}{0.32\textwidth}
        \centering
        \includegraphics[width=\textwidth]{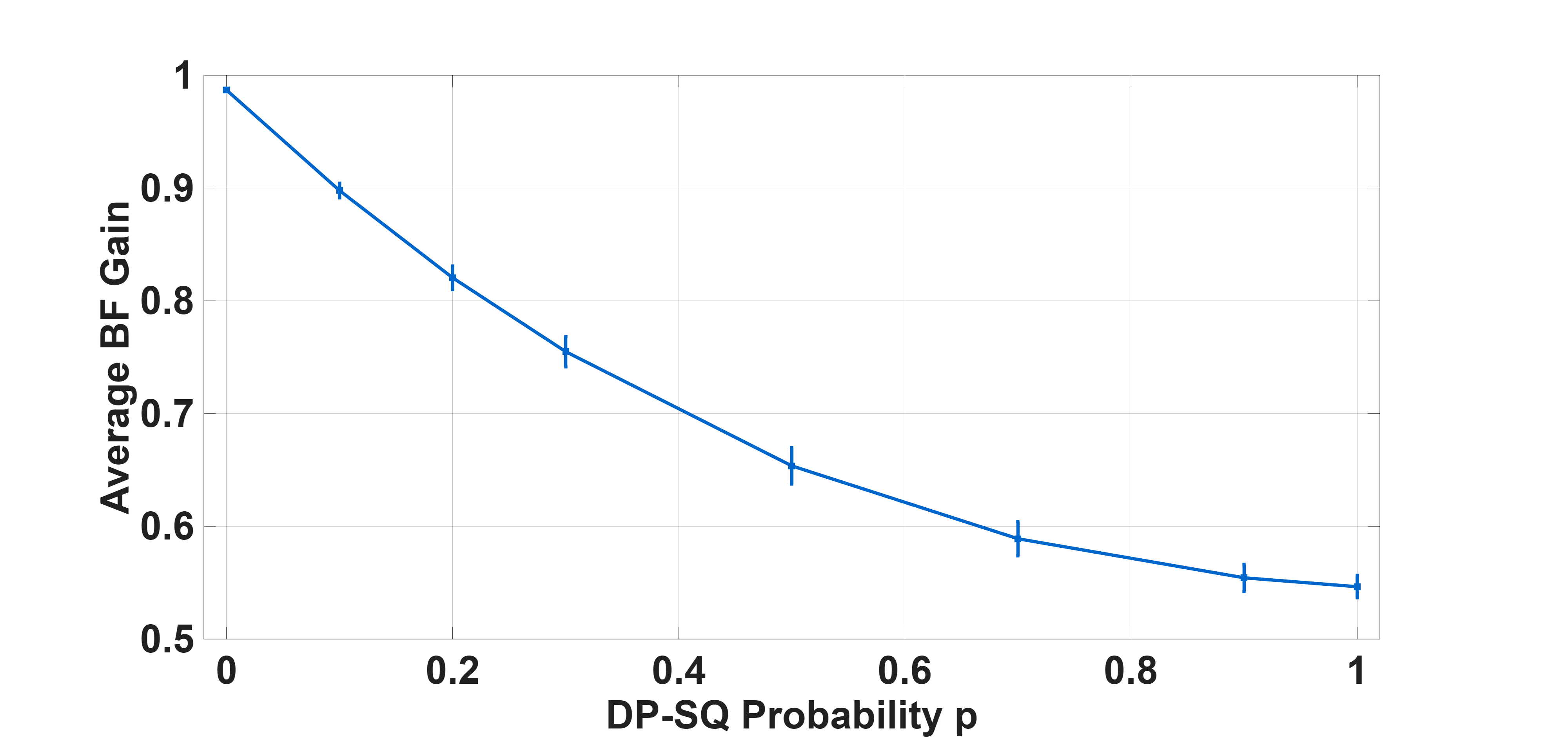}
        \caption{}
        \label{fig:mc_avg_gain}
    \end{subfigure}
    \hfill
    \begin{subfigure}{0.32\textwidth}
        \centering
        \includegraphics[width=\textwidth]{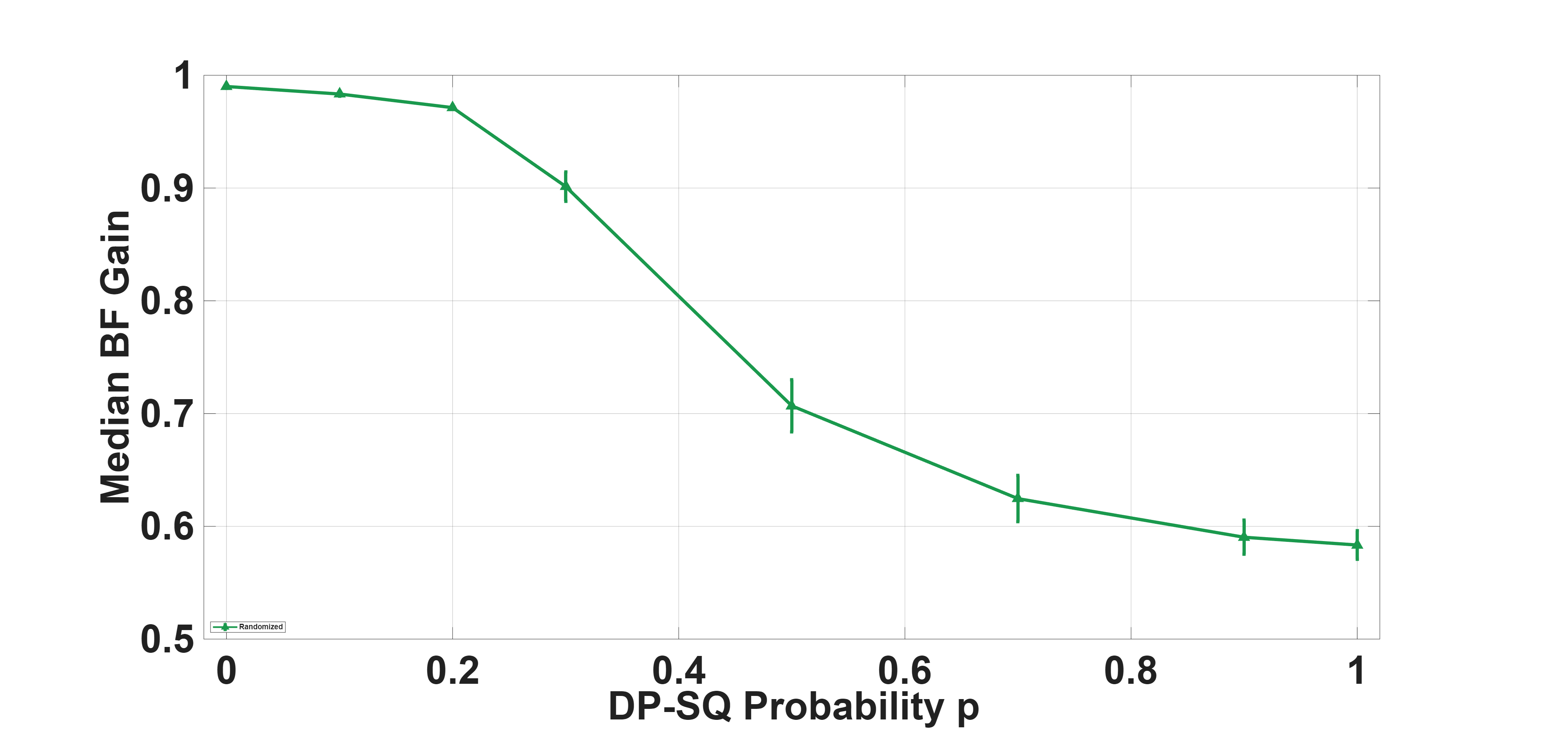}
        \caption{}
        \label{fig:mc_med_gain}
    \end{subfigure}
    \hfill
    \begin{subfigure}{0.32\textwidth}
        \centering
        \includegraphics[width=\textwidth]{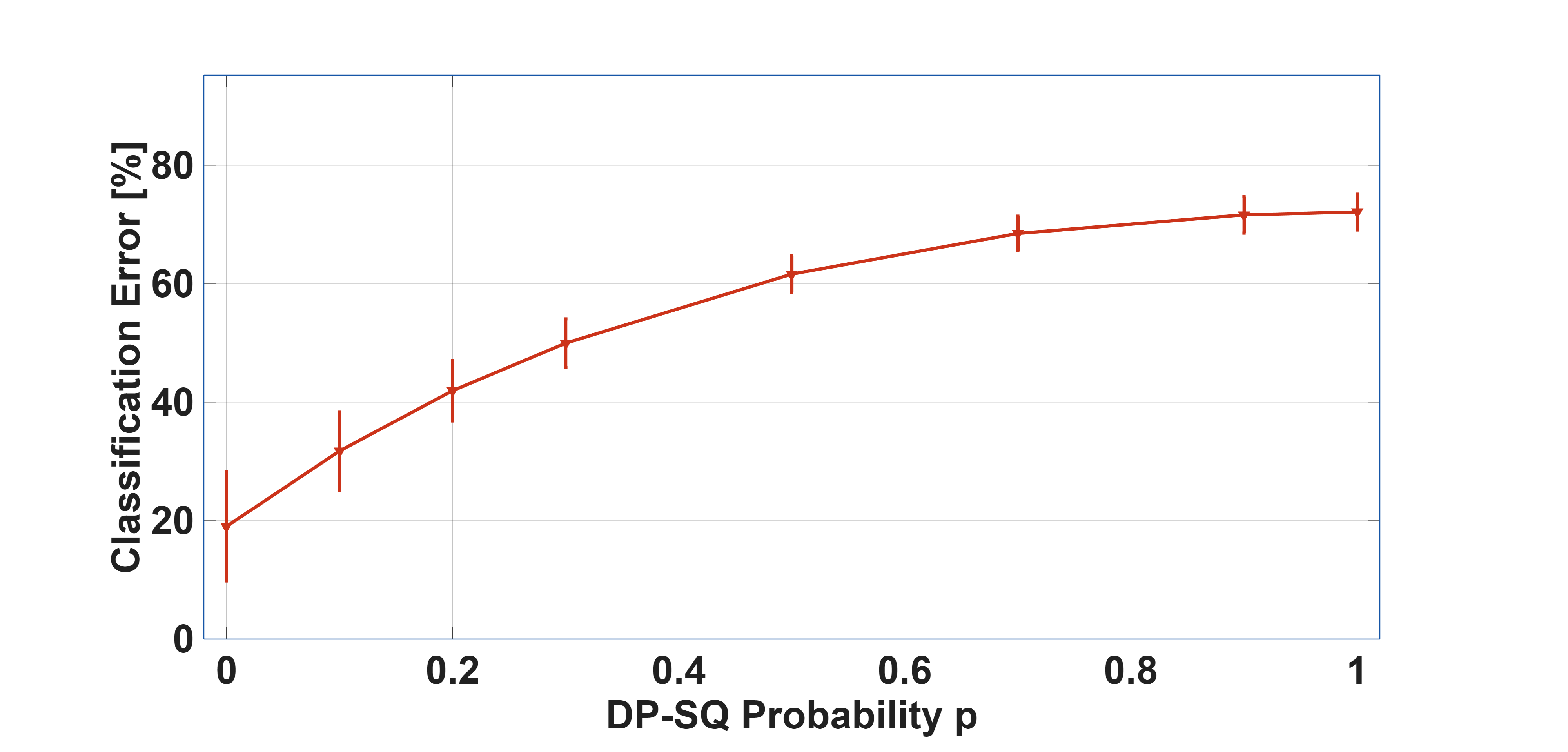}
        \caption{}
        \label{fig:mc_class_err}
    \end{subfigure}

    \caption{\small{Monte Carlo evaluation of the DP-SQ mechanism over $N=1{,}000$
    independent trials with random speed profiles. The dashed red line indicates
    the deterministic baseline ($p=0$). Error bars denote $\pm1\sigma$ across trials.
    (a) Average beamforming gain versus randomization probability $p$.
    (b) Median beamforming gain versus randomization probability $p$.
    (c) Average speed classification error versus randomization probability $p$.}}
    \label{fig:monte_carlo}
\end{figure*}

We further validate this  through a Monte Carlo study comprising 1000 independent trials, each with a randomly generated speed profile drawn from realistic human-motion parameters as defined in Table ~\ref{tab:speed_classification}. Fig.~\ref{fig:monte_carlo}a and Fig.~\ref{fig:monte_carlo}b report
the average and median beamforming gain, respectively, as a function of the randomization probability~$p$. With probability $1-p$ the angle is mapped to its nearest quantization level (deterministic quantization), and with probability $p$ it is replaced by a sample drawn uniformly from the $k$ adjacent candidate levels surrounding the true value. Setting $p = 0$ defaults to the standard scalar quantization with no privacy noise, while $p = 1$ yields maximum randomization (uniform over the $k$-neighborhood). The deterministic baseline ($p=0$) achieves an
average gain of $0.9870$ and a median gain of $0.9900$, both near unity, confirming that standard IEEE~802.11 quantization introduces negligible
beamforming loss. As $p$ increases, both metrics degrade monotonically: at moderate privacy ($p=0.3$) the average and median gains remain above $0.75$ and
$0.90$, respectively, whereas at full randomization ($p=1$) they drop to approximately $0.54$ and $0.58$. Fig.~\ref{fig:monte_carlo}c
presents the average speed-classification error of a micro-Doppler adversary attempting to infer the user's activity zone from the quantized CSI feedback.
At the deterministic baseline the adversary mis-classifies only $19.0\%$ of time windows, but the error rises steeply with~$p$: it exceeds $43\%$ at $p=0.2$, surpasses $50\%$ at $p=0.3$, and saturates near $73\%$ for
$p\geq0.9$, approaching the $75\%$ error rate of a uniform random guess over the four activity classes. The error bars ($\pm1\sigma$ across trials) are tight throughout, confirming that these trends are statistically robust. Taken together, these results demonstrate that the DP-SQ mechanism offers a practical operating region around $p\in[0.1,\,0.3]$ where the adversary's classification ability is substantially degraded (error $>30\%$) while the beamforming gain remains above $75\%$ of the ideal value.

This indicates that the DP mechanism preserves most of the communication utility. At the same time, the same perturbations substantially degrade the eavesdropper’s ability to track the user’s motion, leading to visibly biased and inaccurate velocity trajectories compared to the ground truth. This illustrates the key privacy–utility tradeoff achieved by our design: the beamformer retains near-optimal precoding performance from the CSI privatized feedback, while an adversary observing the perturbed CSI suffers a pronounced loss in speed-estimation accuracy from the reduced information leakage.

It is worth emphasizing that the privacy guarantee of the proposed mechanism is 
\emph{per-shot}: each CSI report is individually protected under 
$\varepsilon$-DP. In our setup, the adversary obtains one CSI measurement 
every $10^{-3}$ seconds over a total communication duration of $T = 10$ seconds, 
resulting in $10{,}000$ DP-protected releases. By applying the advanced 
composition theorem of DP~\cite{dwork2014algorithmic}, the cumulative 
privacy loss grows sublinearly, yielding an overall leakage on the order of
$\varepsilon_{\mathrm{total}} = {O} (\sqrt{T}\, \cdot \max \{\varepsilon_{\psi}, \varepsilon_{\phi} \})$,
demonstrating that even over long communication intervals, the aggregated 
DP leakage remains well controlled.

\begin{table}[t]
\centering
\caption{\small{Speed-based activity classification zones used for the micro-Doppler privacy attack.}}
\label{tab:speed_classification}
\begin{tabular}{c c c}
\hline
\textbf{Zone} & \textbf{Speed Range (m/s)} & \textbf{Activity Label} \\
\hline
1 & $0 \leq |v| < 0.5$   & Stationary \\
2 & $0.5 \leq |v| < 2.5$ & Walking    \\
3 & $2.5 \leq |v| < 5.0$ & Jogging    \\
4 & $5.0 \leq |v| < \infty$ & Running \\
\hline
\end{tabular}
\end{table}

\begin{figure}[t]
    \centering

    \begin{subfigure}{\columnwidth}
        \centering
        \includegraphics[width= 0.85 \linewidth]{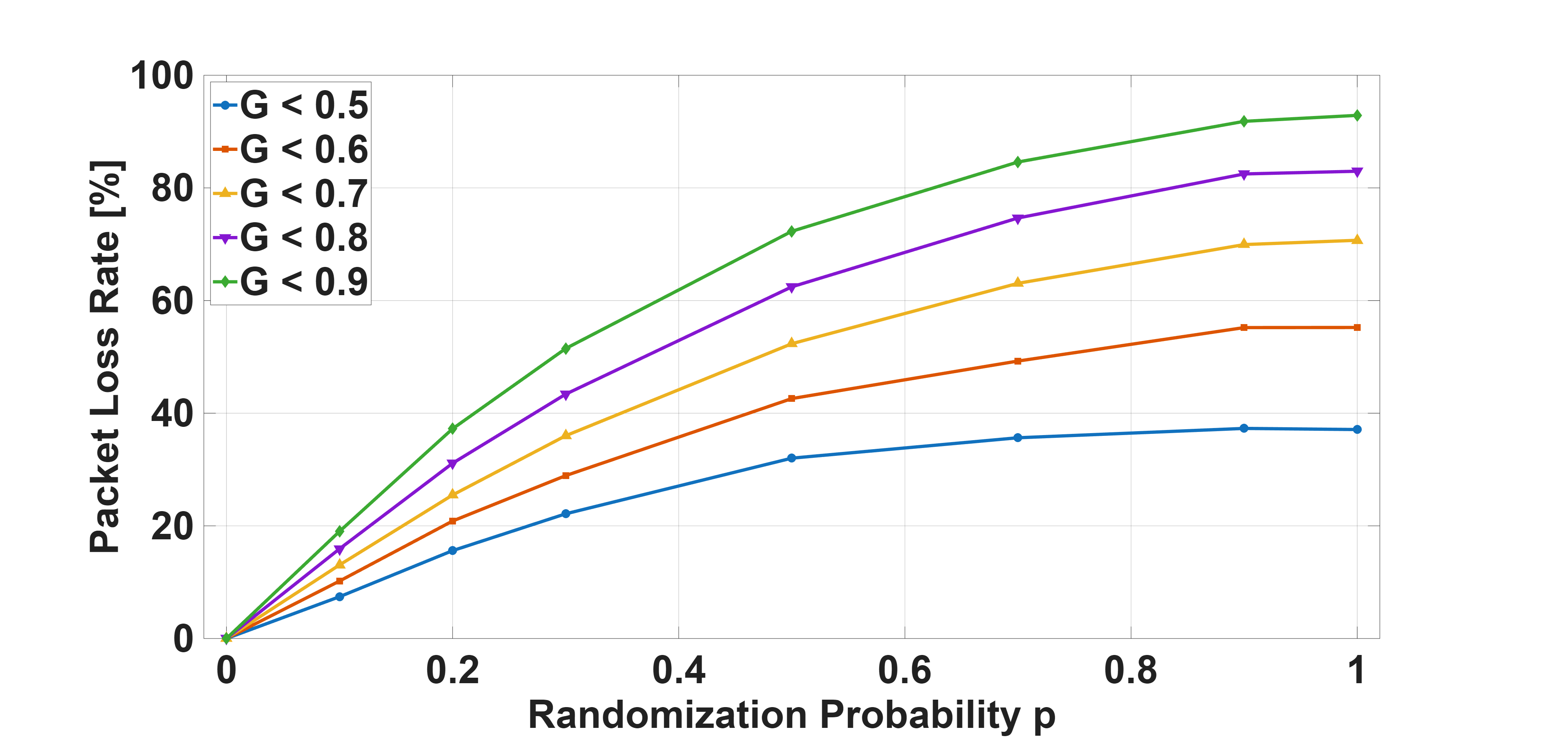}
        \caption{\small{CSI from Deterministic Quantization}}
        \label{fig:Packet_Loss_vs_G}
    \end{subfigure}
    \hfill

    \vspace{6pt}

    \begin{subfigure}{\columnwidth}
        \centering
        \includegraphics[width= 0.85 \linewidth]{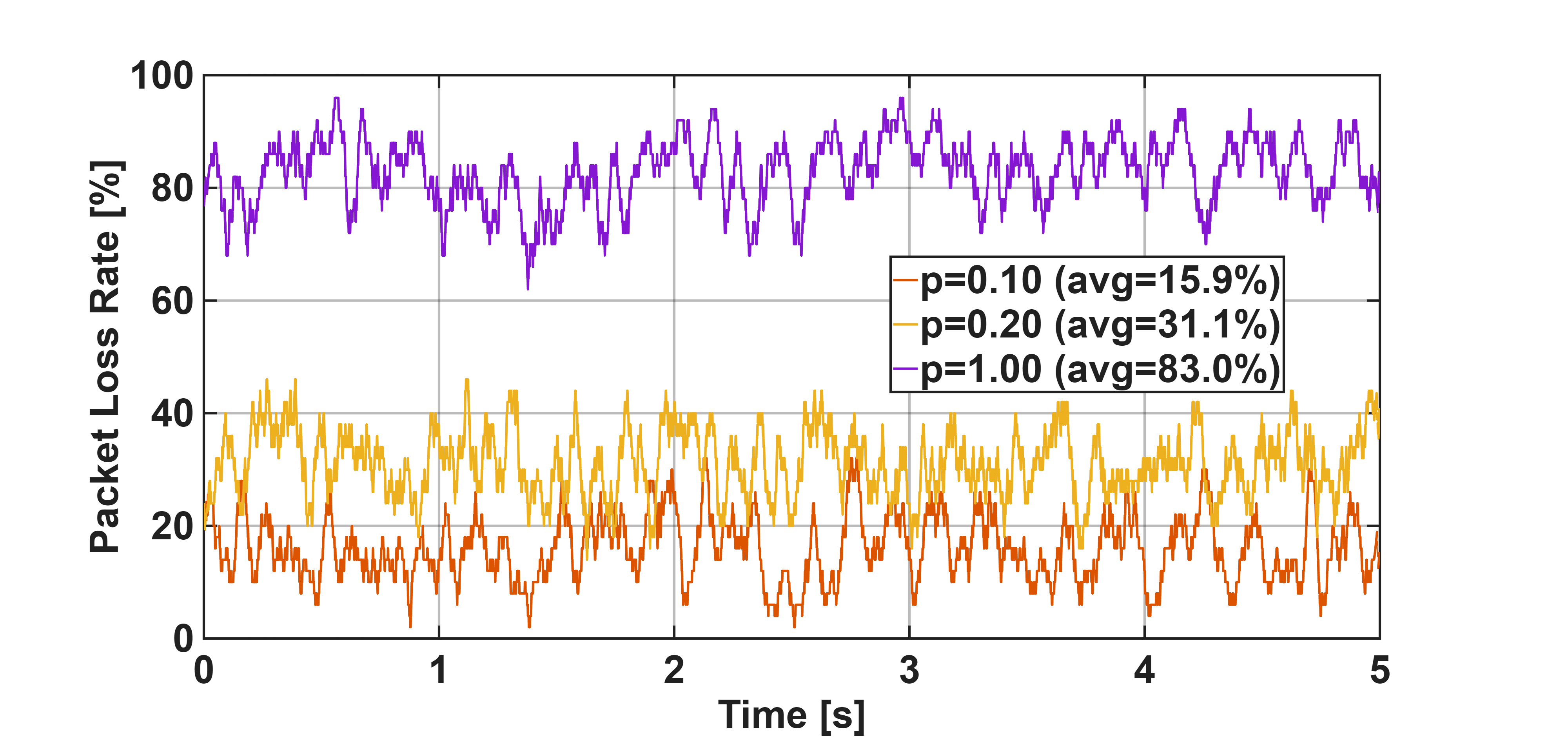}
        \caption{\small{CSI from DP–SQ }}
        \label{fig:Packet_Loss_vs_time}
    \end{subfigure}
    \hfill

    \caption{
        \small{
        Beamformed CSI amplitude across subcarrier index and time for a user transitioning through stationary, walking, jogging, and running activity zones.
        }
    }
    \label{fig:packet-loss}
\end{figure}

\noindent \textbf{Impact on Packet Loss.}  Fig.~\ref{fig:packet-loss}a shows the packet loss rate as a function of the randomization probability $p$ for beamforming various gain thresholds,
$G_{\min}$. All curves originate near zero at $p = 0$ and increase monotonically, confirming that the communication cost of DP-SQ scales smoothly with the
privacy level. At the most stringent threshold ($G_{\min} = 0.9$), the loss rate rises steeply, reaching approximately $80\%$ at $p = 0.7$ and exceeding $90\%$ at $p = 1.0$. Relaxing the threshold to $G_{\min} = 0.8$ reduces the loss to roughly $70\%$ at full randomization, while the $G_{\min} = 0.5$ curve remains below $40\%$ even at $p = 1.0$, indicating that links operating with comfortable SNR margins can tolerate aggressive randomization. Fig.~\ref{fig:packet-loss}b presents the time-resolved packet loss for the $G_{\min} = 0.8$ threshold as the randomization probability increases.  Notably, the loss events do not cluster into prolonged outages but
instead appear uniformly distributed in time, suggesting that DP-SQ
degrades individual snapshot gains rather than inducing correlated
fading bursts.

\section{Conclusion \& Future Work}
\label{sec:conclusion}

In this paper, we have presented a standards-compatible framework for private CSI feedback
based on DP quantization of the Givens rotation
and phase angles used in IEEE~802.11ac/ax compressed beamforming.
By replacing deterministic angle quantization with an $\varepsilon$-DP
stochastic mechanism, the proposed design preserves the existing
feedback format while providing formal privacy guarantees against
activity, identity, and environment inference attacks. Numerical simulations under realistic 802.11 configurations demonstrate
that DP angle quantization significantly suppresses adversarial
classification accuracy while incurring only modest beamforming loss at
the access point.
These results highlight that meaningful privacy can be delivered within
the current 802.11 compressed feedback architecture without modifying
frame formats, pilot structures, or channel sounding procedures. The same solution can in principle be also applied to other 802.11 standards that support higher frequency mmWave bands (such as 802.11ad/11ay) or standards that incorporate more advanced hardware. Thus, our approach can be readily adapted to future wireless systems. Thus, future work includes extending the mechanism to multi-user beamforming \cite{he2026beamforming},
hybrid-precoding architectures, RIS-assisted systems \cite{he2026invisible}, and adaptive DP
budget allocation across streaming feedback packets.
Our findings open a new direction in integrating rigorous privacy
mechanisms into commodity Wi-Fi standards to protect users from passive
CSI-based sensing attacks.

\bibliographystyle{IEEEtran}
\bibliography{myreferences}

@book{dwork2014algorithmic,
  title     = {The Algorithmic Foundations of Differential Privacy},
  author    = {Dwork, Cynthia and Roth, Aaron},
  series    = {Foundations and Trends in Theoretical Computer Science},
  volume    = {9},
  number    = {3--4},
  pages     = {211--407},
  publisher = {Now Publishers Inc.},
  year      = {2014}
}

@inproceedings{jiang2021wifi,
  title={Protecting privacy in {WiFi} sensing: A survey and new insights},
  author={Jiang, Wei and Li, Chao and Zhang, Xin},
  booktitle={Proceedings of the 2021 IEEE Conference on Communications and Network Security (CNS)},
  year={2021},
  pages={1--9}
}

@article{wang2017device,
  title={Device-free human activity recognition using commercial {WiFi} devices},
  author={Wang, Wei and Liu, Alex X and Shahzad, Muhammad and Ling, Kang and Lu, Sanglu},
  journal={IEEE Journal on Selected Areas in Communications},
  volume={35},
  number={5},
  pages={1118--1131},
  year={2017},
}

@inproceedings{abdelnasser2015wigest,
  title={{WiGest}: A ubiquitous {WiFi}-based gesture recognition system},
  author={Abdelnasser, Heba and Youssef, Moustafa and Harras, Khaled A},
  booktitle={Proceedings of the 2015 IEEE Conference on Computer Communications (INFOCOM)},
  pages={1472--1480},
  year={2015}
}

@article{park2013ieee80211ac,
  title   = {{IEEE} 802.11ac: Dynamic link adaptation and multi‐user MIMO},
  author  = {Park, Moo‐Ryong and Chen, Hao and Lee, S. and Shin, Kang G.},
  journal = {IEEE Communications Magazine},
  volume  = {51},
  number  = {10},
  pages   = {90--96},
  year    = {2013},
  doi     = {10.1109/MCOM.2013.6619577}
}

@inproceedings{liu2024lendmeyourbeam,
  title     = {Lend Me Your Beam: Privacy Implications of Plaintext Beamforming Feedback in {WiFi}},
  author    = {Liu, Yuqi and Zeng, Yuanjie and Uluagac, A. Selcuk and Jana, Suman},
  booktitle = {Proceedings of the Network and Distributed System Security Symposium (NDSS)},
  year      = {2024}
}

@article{cominelli2024physical,
  author={Cominelli, Marco and Shahcheraghi, Shaghayegh and Link, Jakob and Hollick, Matthias and Cerutti, Federico and Gringoli, Francesco and Asadi, Arash},
  journal={IEEE Transactions on Wireless Communications}, 
  title={Physical-Layer Privacy via Randomized Beamforming Against Adversarial {Wi-Fi} Sensing: Analysis, Implementation, and Evaluation}, 
  year={2024},
  volume={23},
  number={12},
  pages={19603-19617},
}

@inproceedings{shenoy2022rf,
  title={{RF}-Protect: privacy against device-free human tracking},
  author={Shenoy, Jayanth and Liu, Zikun and Tao, Bill and Kabelac, Zachary and Vasisht, Deepak},
  booktitle={Proceedings of the ACM SIGCOMM 2022 Conference},
  pages={588--600},
  year={2022}
}

@standard{ieee80211n2012,
  title        = {{IEEE} Standard for Information Technology--Telecommunications and Information Exchange between Systems Local and Metropolitan Area Networks--Specific requirements--Part 11: Wireless LAN Medium Access Control {(MAC)} and Physical Layer {(PHY)} Specifications Amendment 3: Enhancements for Higher Throughput},
  organization = {IEEE},
  number       = {802.11n-2012},
  year         = {2012},
  doi          = {10.1109/IEEESTD.2012.6209369},
  note         = {{I}EEE Std 802.11n-2012}
}

@book{goldsmith2005wireless,
  title={Wireless Communications},
  author={Goldsmith, Andrea},
  year={2005},
  publisher={Cambridge University Press}
}

@article{zhu2025csi,
  title={{CSI}-Bench: A Large-Scale In-the-Wild Dataset for Multi-task {WiFi} Sensing},
  author={Zhu, Guozhen and Hu, Yuqian and Gao, Weihang and Wang, Wei-Hsiang and Wang, Beibei and Liu, KJ},
  journal={arXiv preprint arXiv:2505.21866},
  year={2025}
}

@book{perahia2013next,
  title={Next {G}eneration {W}ireless {L}ANs: 802.11 n and 802.11 ac},
  author={Perahia, Eldad and Stacey, Robert},
  year={2013},
  publisher={Cambridge University Press}
}

@article{geraci2025wi,
  title={{Wi-Fi}: Twenty-five years and counting},
  author={Geraci, Giovanni and Meneghello, Francesca and Wilhelmi, Francesc and Lopez-Perez, David and Val, I{\~n}aki and Giordano, Lorenzo Galati and Cordeiro, Carlos and Ghosh, Monisha and Knightly, Edward and Bellalta, Boris},
  journal={arXiv preprint arXiv:2507.09613},
  year={2025}
}

@article{he2026invisible,
  title={Invisible Walls: Privacy-Preserving ISAC Empowered by Reconfigurable Intelligent Surfaces},
  author={He, Yinghui and Fan, Long and Xie, Lei and Niyato, Dusit and Yuen, Chau and Luo, Jun},
  journal={arXiv preprint arXiv:2601.04488},
  year={2026}
}

@book{el2011network,
  title={Network information theory},
  author={El Gamal, Abbas and Kim, Young-Han},
  year={2011},
  publisher={Cambridge university press}
}

@article{he2026beamforming,
  title={Beamforming-Enabled Integrated Sensing and Communication over Commodity Multi-User Wi-Fi},
  author={He, Yinghui and Xu, Mingming and Chen, Zhe and Xiao, Fu and Luo, Jun},
  journal={IEEE Transactions on Mobile Computing},
  year={2026}
}

\appendices
\section*{Auxilary Lemma}
\label{app:givens_dp_sq}


\begin{lemma}[Stage-wise Perturbation Bound for Givens Cascades]
\label{lem:stagewise_perturbation}
Consider a unit-norm vector $\mathbf v_i^{(m)} \in \mathbb C^{N_t}$ generated through the Annex--N Givens parameterization. Let $\mathbf v_i^{(m-1)}$ and $\mathbf v_i^{(m)}$ denote the vectors immediately before and after perturbing the $m$-th pair of angles $(\psi_{i,m},\phi_{i,m})$ by $(\Delta\psi_{i,m},\Delta\phi_{i,m})$, respectively. Then,
\[
\|\mathbf v_i^{(m)}-\mathbf v_i^{(m-1)}\|_2^2
\le
8\!  \left(
\sin^2\tfrac{\Delta\psi_{i,m}}{2}
+\sin^2\!\Psi_{i,+}\,\sin^2\tfrac{\Delta\phi_{i,m}}{2}
\right).
\]
where $\Psi_{i,+}$ denotes the cumulative mixing angle determining the energy of the branch affected by the phase perturbation.
\end{lemma}

\begin{proof}
At stage $m$, only a $2\times2$ block of the Givens cascade is modified. Let $\mathbf u \in \mathbb C^2$ denote the corresponding sub-vector with $\|\mathbf u\|_2 \le 1$, and let $\mathbf B(\psi,\phi)$ denote the local transformation. Then
\begin{align}
\mathbf z(\psi,\phi) = \mathbf B(\psi,\phi)\mathbf u, 
\qquad
\widetilde{\mathbf z}
=
\mathbf B(\psi+\Delta\psi,\phi+\Delta\phi)\mathbf u. \nonumber
\end{align}

We next isolate the effect of angle perturbations at each stage by separating the mixing and phase contributions. Using the decomposition
\begin{align}
\widetilde{\mathbf z}-\mathbf z
=
\mathbf e_\psi+\mathbf e_\phi,
\nonumber
\end{align}
where
\begin{align}
\mathbf e_\psi
&:=
\big(\mathbf B(\psi+\Delta\psi,\phi+\Delta\phi)-\mathbf B(\psi,\phi+\Delta\phi)\big)\mathbf u, \nonumber \\
\mathbf e_\phi
&:=
\big(\mathbf B(\psi,\phi+\Delta\phi)-\mathbf B(\psi,\phi)\big)\mathbf u, \nonumber 
\end{align}
and applying $\|x+y\|_2^2\le 2\|x\|_2^2+2\|y\|_2^2$, we obtain
\begin{align}
\|\widetilde{\mathbf z}-\mathbf z\|_2^2
\le
2\|\mathbf e_\psi\|_2^2+2\|\mathbf e_\phi\|_2^2. 
\label{eq:split_terms}
\end{align}

\noindent \textbf{Mixing angle perturbation.} The term $\mathbf e_\psi$ corresponds to perturbing only the Givens rotation. Since multiplication by a unit-modulus phase does not affect the norm,
\begin{align}
\|\mathbf e_\psi\|_2
\le
\|\mathbf R(\psi+\Delta\psi)-\mathbf R(\psi)\|_2 \,\|\mathbf u\|_2. 
\nonumber 
\end{align}
Using $\|\mathbf u\|_2\le 1$ and the identity
\begin{align}
\|\mathbf R(\psi+\Delta\psi)-\mathbf R(\psi)\|_2
=
2\left|\sin\frac{\Delta\psi}{2}\right|,
\nonumber 
\end{align}
we obtain
\begin{equation}
\|\mathbf e_\psi\|_2^2
\le
4\sin^2\frac{\Delta\psi}{2}.
\label{eq:mixing_final}
\end{equation}

\noindent \textbf{Phase perturbation.} The term $\mathbf e_\phi$ corresponds to perturbing only the phase. For any complex scalar $w$,
\begin{align}
|e^{j(\phi+\Delta\phi)}w-e^{j\phi}w|^2
=
|e^{j\Delta\phi}-1|^2\,|w|^2
=
4|w|^2\sin^2\frac{\Delta\phi}{2}. \nonumber 
\end{align}
Under the Annex--N parameterization, the affected branch satisfies
\begin{align}
|w|^2 \le \sin^2 \Psi_{i,+}, 
\nonumber
\end{align}
which yields
\begin{equation}
\|\mathbf e_\phi\|_2^2
\le
4\sin^2\Psi_{i,+}\,\sin^2\frac{\Delta\phi}{2}.
\label{eq:phase_final}
\end{equation}

Substituting \eqref{eq:mixing_final} and \eqref{eq:phase_final} into \eqref{eq:split_terms}, we obtain the upper bound presented in Lemma~\ref{lem:stagewise_perturbation}, which completes the proof.

\end{proof}

\end{document}